\documentclass[runningheads]{llncs}
\usepackage[T1]{fontenc}
\usepackage{graphicx}
\usepackage{comment}

\usepackage{xcolor}
\usepackage{hyperref}

\usepackage[normalem]{ulem}

\usepackage{amsmath} \usepackage{amssymb} \usepackage{latexsym}
\usepackage{stmaryrd}
\usepackage[notextcomp]{stix}
\usepackage{threeparttable}
\usepackage{makecell}
\usepackage{subcaption}
\newcommand{\vfi}{{\varphi}}
\newcommand{\ar}{{\Rightarrow}}
\newcommand{\btop}{\boldsymbol{\top}}
\newcommand{\bbot}{\boldsymbol{\bot}}
\newcommand{\band}{\boldsymbol{\wedge}}
\newcommand{\bor}{\boldsymbol{\vee}}
\newcommand{\bneg}{\boldsymbol{\neg}}
\newcommand{\cmpl}{{-}}
\newcommand{\bcmpl}{{\boldsymbol{-}}}

\newcommand{\imp}{\longrightarrow}
\newcommand{\limp}{\rightarrow}
\newcommand{\bimp}{\boldsymbol{\rightarrow}}
\newcommand{\dimp}{\longleftrightarrow}

\newcommand{\lstar}{\bigwhitestar}
\newcommand{\lbstar}{\bigstar}

\newcommand{\bdimp}{\boldsymbol{\leftrightarrow}}
\newcommand{\bdiff}{\boldsymbol{\leftharpoonup}}
\newcommand{\bsdiff}{\boldsymbol{\vartriangle}}

\newcommand{\bforall}{\boldsymbol{\forall}}
\newcommand{\bexists}{\boldsymbol{\exists}}
\newcommand{\bapprox}{\boldsymbol{=}}
\newcommand{\bprec}{\boldsymbol{\leq}}
\newcommand{\bsucc}{\boldsymbol{\geq}}

\newcommand{\bsup}{\boldsymbol{\bigvee}}
\newcommand{\binf}{\boldsymbol{\bigwedge}}
\newcommand{\bcirc}{\boldsymbol{\circ}}
\newcommand{\bbcirc}{\boldsymbol{\bullet}}
\newcommand{\bstar}{\boldsymbol{\bigwhitestar}}
\newcommand{\bbstar}{\boldsymbol{\bigstar}}

\newcommand{\bsqand}{\ddot{\band}}
\newcommand{\bsqor}{\ddot{\bor}}
\newcommand{\bsqimp}{\ddot{\bimp}}

\newcommand{\bsqtop}{\ddot{\btop}}
\newcommand{\bsqbot}{\ddot{\bbot}}

\newcommand{\bPi}{\boldsymbol{\Pi}}
\newcommand{\bSigma}{\boldsymbol{\Sigma}}
\newcommand{\bn}{\boldsymbol{n}}
\newcommand{\be}{\boldsymbol{e}}

\newcommand{\qdash}{\,{\dashv}{\vdash}\,}

\newcommand{\bool}{o}

\newcommand{\itype}{\alpha} 
\newcommand{\equdef}{\eqdef}
\newcommand{\equ}{=}
\newcommand{\ww}{\omega}

\newcommand{\wsig}{(\ww)\sigma}

\newcommand{\A}{\mathfrak{A}}

\newcommand{\I}{\mathcal{I}}
\newcommand{\C}{\mathcal{C}}
\newcommand{\B}{\mathcal{B}}
\newcommand{\F}{\mathcal{F}}
\newcommand{\D}{\mathcal{D}}
\newcommand{\R}{\mathcal{R}}

\newcommand{\E}{\mathcal{E}}
\newcommand{\Q}{\mathcal{Q}}

\newcommand{\Lang}{\mathcal{L}}

\newcommand{\T}{\mathcal{T}}
\newcommand{\V}{\mathcal{V}}

\newcommand{\Const}{\C\textit{onst}}
\newcommand{\Var}{\V\textit{ar}}

\newcommand{\MONO}{\texttt{MONO}}
\newcommand{\ADDI}{\texttt{ADDI}}
\newcommand{\iADDI}{\texttt{iADDI}}
\newcommand{\MULT}{\texttt{MULT}}
\newcommand{\iMULT}{\texttt{iMULT}}
\newcommand{\EXPN}{\texttt{EXPN}}
\newcommand{\CNTR}{\texttt{CNTR}}
\newcommand{\NORM}{\texttt{NORM}}
\newcommand{\DNRM}{\texttt{DNRM}}
\newcommand{\IDEM}{\texttt{IDEM}}

\newcommand{\meetcl}{\texttt{meet\_closed}}
\newcommand{\joincl}{\texttt{join\_closed}}
\newcommand{\supmcl}{\texttt{supremum\_closed}}
\newcommand{\infmcl}{\texttt{infimum\_closed}}

\newcommand{\cmplbar}{{\--}} 

\newcommand{\fpbar}{{\overline{\texttt{fp}}}}
\newcommand{\dbar}{{\overline{\texttt{d}}}}

\newcommand{\fp}[1]{\texttt{fp}~{#1}}
\newcommand{\supfp}[1]{{#1}^{\texttt{fp}}}
\newcommand{\supcfp}[1]{{#1}^{\fpbar}}
\newcommand{\supc}[1]{{#1}^{\cmplbar}}
\newcommand{\supd}[1]{{#1}^{\texttt{d}}}
\newcommand{\supdc}[1]{{#1}^{\dbar}}

\begin{document}
\title{Semantical Investigations on Non-classical Logics \\
	with Recovery Operators: Negation
}
%
%
\author{David Fuenmayor\inst{1,2}\orcidID{0000-0002-0042-4538}}
\authorrunning{D.~Fuenmayor}
%
\institute{AI Systems Engineering, Otto-Friedrich-Universit\"at Bamberg, Germany \email{david.fuenmayor@uni-bamberg.de}
\and Mathematics and Computer Science, Freie Universit\"at Berlin, Germany
}
\maketitle              
\begin{abstract}
We investigate mathematical structures that provide natural semantics for families of (quantified) non-classical logics featuring special unary connectives, known as \textit{recovery operators}, that allow us to `recover' the properties of classical logic in a controlled manner. 
These structures are known as \textit{topological Boolean algebras}, which are Boolean algebras extended with additional operations subject to specific conditions of a topological nature.
In this study we focus on the paradigmatic case of negation. We demonstrate how these algebras are well-suited to provide a semantics for some families of paraconsistent \textit{Logics of Formal Inconsistency} and paracomplete \textit{Logics of Formal Undeterminedness}. These logics feature recovery operators used to earmark propositions that behave `classically' when interacting with non-classical negations.
Unlike traditional semantical investigations, which are carried out in natural language (extended with mathematical shorthand), our formal meta-language is a system of higher-order logic (HOL) for which automated reasoning tools exist. In our approach, topological Boolean algebras are encoded as algebras of sets via their Stone-type representation. We use our higher-order meta-logic to 
define and interrelate several transformations on unary set operations, which naturally give rise to a \textit{topological cube of opposition}.
Additionally, our approach enables a uniform characterization of propositional, first-order, and higher-order quantification, including restrictions to constant and varying domains. With this work, we aim to make a case for the utilization of automated theorem proving technology for conducting computer-supported research in non-classical logics. All the results presented in this paper have been formally verified, and in many cases obtained, using the Isabelle/HOL proof assistant.

\end{abstract}

\pagebreak

\section{Introduction}

From the perspective of computer science and artificial intelligence (e.g.,~knowledge representation) it is natural to seek `non-classical' logical formalisms that accommodate reasoning under real-world conditions, such as the presence of partial or contradictory information.
For instance, by adopting an epistemic approach towards paraconsistency and paracompleteness in logic (e.g.,~\cite{CR2017}), we can view these as coping mechanisms for the inevitable occurrence of `epistemically suboptimal' situations, where cognizers might encounter incomplete and/or contradictory evidence regarding the truth or falsity of propositions. This epistemic approach underscores the need to augment a logic with expressive means to qualify the (quality of) evidence for individual propositions, and to discern which among them can be designated `safe' for classical reasoning. Several researchers suggestively refer to object-logical connectives fulfilling this role as \textit{recovery operators} \cite{CCR2020}. These are paradigmatically instantiated in the families of paraconsistent
\textit{Logics of Formal Inconsistency} (LFIs; introduced in \cite{CM}, cf.~also \cite{BookCC16,CCR2020}) and paracomplete \textit{Logics of Formal Undeterminedness} (LFUs; introduced in \cite{Marcos2005}, cf.~also \cite{CCR2020}).

On a related note, spurred by ongoing efforts to implement symbolic capabilities in AI systems, particularly with respect to automated legal and normative reasoning (see e.g.~\cite{J48,ECAI} and references therein), a specific technique, known as \textit{shallow semantical embedding} \cite{J41,J23}, has been developed with the intent of repurposing automated theorem proving technology for classical first- and higher-order logic to provide \textit{practically effective}%
\footnote{Effective from a pragmatic, best-effort AI engineering perspective, meaning performing adequately in the challenging (predominantly \textit{undecidable}) tasks encountered in contemporary AI. While theoretical computability and complexity issues can significantly influence the design of certain safety-critical systems (e.g., in cybersecurity), they cannot guarantee (nor deter) practical success of automated reasoning tools. A good example of this is the impressive performance of SAT-solvers (and more recently SMT-solvers) in formal verification, where they routinely solve complex problems (e.g., involving hundreds of variables) otherwise deemed `intractable' by (worst-case) theoretical analysis.}
automation for families of \textit{quantified} non-classical logics. This technique introduces families of \textit{object logics} by encoding their connectives as terms of a fully formal \textit{meta-logic}, in which their truth conditions are stated and semantical constraints are axiomatized. This approach aligns closely with the original idea as proposed by Tarski \cite{Tarski1933}, and which has influenced several generations of logicians, for example, by motivating the development of model theory. In contrast to traditional model-theoretical investigations that use natural language (extended with mathematical shorthand) as a meta-language, the meta-logic used in shallow semantical embeddings is a system of higher-order logic (HOL), extending Church's simple theory of types \cite{Church40,SEPTT}, which lies at the heart of many modern automated theorem provers and interactive proof assistant systems \cite{B5}. This approach has been utilized for interactive (meta-)theoretical investigations in (and about) quantified modal and non-classical logics using mathematical proof assistants, in particular Isabelle/HOL \cite{Isabelle}. Moreover, this research has been applied to the logico-pluralistic, computer-supported formalization and analysis of philosophical arguments and normative theories (cf.~\cite{ECAI,J48,J41} and references therein).

Building upon prior research on shallow semantical embeddings, we present in this work a class of algebras aimed at providing natural semantics for families of (quantified) non-classical logics featuring recovery operators.
These structures are referred to as \textit{topological Boolean algebras} (TBAs).\footnote{We were inspired to choose this term by Rasiowa \& Sikorski's monograph \textit{The Mathematics of Metamathematics} \cite{TMMT}. McKinsey \& Tarski's \textit{The Algebra of Topology} \cite{AOT} can be considered the seminal work investigating this kind of structures, where they are termed \textit{closure algebras}.} TBAs are Boolean algebras extended with additional operations that satisfy axiomatic conditions of a topological nature. In this study we focus on the unary case and refer to these additional unary operations as \textit{operators}. Traditional examples of operators in logic thus include (unary) modalities and negations.
Furthermore, we concentrate on the paradigmatic case of negation (and corresponding recovery operators). We show how TBAs are aptly suited to provide semantics for certain families of (paraconsistent) LFIs and (paracomplete) LFUs.

We adopt a natural approach towards the shallow semantical embedding of TBAs, in which they become encoded as algebras of sets (via their Stone-type representation). We utilize our higher-order meta-logic to 
define and interrelate certain special transformations on operators, such that they naturally give rise to a commutative diagram in the form of a \textit{topological cube of opposition}.
We also discuss how our approach enables a uniform characterization of propositional, first-order, and higher-order quantification, including restrictions to constant and varying domains.

With this work we also want to make a case for the utilization of formal meta-languages and automated reasoners for doing computer-supported research in non-classical logics. In fact, all presented results have been formally verified, and in many cases obtained, using the Isabelle/HOL proof assistant \cite{Isabelle}.
Readers interested in examining the Isabelle/HOL sources corresponding to the results discussed in this paper are encouraged to visit our GitHub repository\footnote{Visit \url{https://github.com/davfuenmayor/topological-semantics}. This repository hosts the development version of the corresponding \textit{Archive of Formal Proofs} \cite{AFP} entry.} where they can access the most recent version of our ongoing formalization work. Many of the results discussed here were previously presented at the 3rd International Conference on Non-Classical Modal and Predicate Logics (NCMPL 2021).

In the following section \S\ref{sec:preliminaries} we have made an effort to present some relevant background material in an uniform manner. In particular, much of the material on TBAs has been gathered from various sources scattered throughout the literature.\footnote{The early work on the axiomatic foundations of topology by Kuratowski \cite{Kuratowski1,Kuratowski2} and Zarycki \cite{Zarycki1,Zarycki2,Zarycki3} deserves special mention here. We find it rather unfortunate, especially for logicians, that	this way of introducing topology has not received the recognition it deserves. The interested reader can now find English translations of Zarycki's works which have been recently uploaded on the web (thanks to Mark Bowron). See \url{https://www.researchgate.net/scientific-contributions/Miron-Zarycki-2016157096}.} This material will help to establish the required conceptual framework for the discussion in \S\ref{sec:investigations}, where we use the shallow semantical embedding approach to develop a theory of unary operations (operators) on top of Boolean algebras encoded as algebras of sets. We use this theory to provide uniform semantics for families of \textit{quantified} LFIs and LFUs via TBAs. We conclude the paper in \S\ref{sec:conclusion}.

\section{Conceptual Preliminaries}\label{sec:preliminaries}
\subsection{Paraconsistent and Paracomplete Logics}\label{subsec:prelim-negation}

Logics featuring non-classical negation-like operators are not only of philosophical interest but also have important applications in computer science and artificial intelligence. In particular, they aid in the development of robust knowledge representation and reasoning techniques that can operate under conditions of incomplete and contradictory information.
 
A logic is termed \textit{paraconsistent} when it features a negation operator ($\bneg$) for which the principle of \textit{ex contradictione (sequitur) quodlibet} (ECQ), also colloquially referred to as `explosion', does not hold: not everything follows from a contradiction $A \land \neg A$.
Dually, paracomplete logics feature a negation for which the principle of \textit{tertium non datur} (TND), aka.~`excluded middle', is not valid: formulas of the form $A \lor \neg A$ are not tautologies. Intuitionistic logic, for instance, is a well-known paracomplete logic (out of infinitely many) and dual-intuitionistic logic is paraconsistent.
Broadly speaking, paraconsistent logics can be said to `tolerate contradictions', while paracomplete logics do not `require exhaustiveness' of available information or evidence.\footnote{Note that we assume an epistemic interpretation of paraconsistency and paracompleteness (as in e.g.~\cite{CR2017}). This is justified by our ongoing efforts in applying non-classical logics in areas of AI like knowledge representation and reasoning \cite{ECAI,J48}.}

It is known that any unary operation that validates both ECQ and TND (wrt.~the Boolean $\land$ \& $\lor$ connectives) collapses into classical negation. Thus, whenever we require a `weaker' negation (upon a Boolean base), which invalidates some logical property of its classical counterpart, we are compelled to relinquish either TND or ECQ (or sometimes both). Consequently, our logic will inevitably become either paracomplete or paraconsistent (or sometimes both).\footnote{Logics that are both paraconsistent and paracomplete are referred to as \textit{paradefinite} (or \textit{paranormal}) in the literature. We direct the reader to \cite{BookCC16} for a comprehensive discussion on the theory and applications of paraconsistent and other non-classical logics.}

\begin{figure}
	\centering
			\includegraphics[width=.48\textwidth]{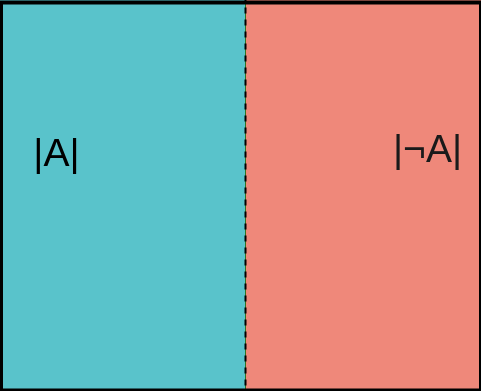}
		\caption{Suggestive pictorial representation of classical negation.}
		\label{fig:neg-classical}
\end{figure}

\begin{figure}
	\centering
	\begin{subfigure}{0.49\textwidth}
		\centering
		\includegraphics[width=.90\textwidth]{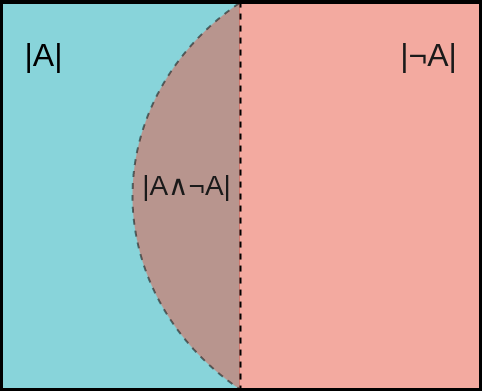}		
		\caption{Paraconsistency gives rise to \textit{truth-gluts}}
		\label{fig:neg-glut}
	\end{subfigure}\vspace{5pt}
	\begin{subfigure}{0.49\textwidth}
		\centering
		\includegraphics[width=.90\textwidth]{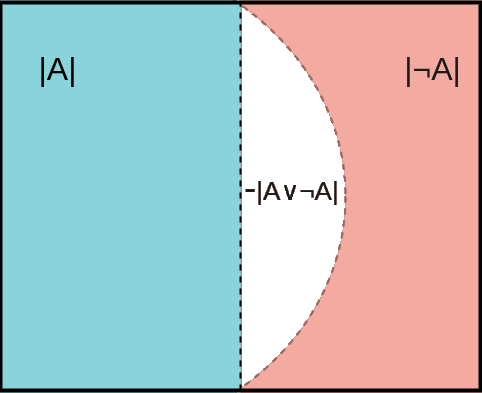}		
		\caption{Paracompleteness gives rise to \textit{truth-gaps}}
		\label{fig:neg-gap}
	\end{subfigure}\hfill%
	\caption{Suggestive pictorial representation of non-classical negations.}
	\label{fig:neg-nonclassical}
\end{figure}

Fig.~\ref{fig:neg-classical} shows a suggestive pictorial representation of the well-known behavior of the classical negation, where the notation $|A|$ represents the semantical denotation of some sentence $A$ as its so-called `truth-set', i.e., the set of situations (aka.~`worlds') in which $A$ happens to be true. As illustrated, classical logic makes the assumption that we always have complete and non-contradictory information or evidence as to whether $A$ or its negation $\neg A$ is the case.
It is instructive to compare Fig.~\ref{fig:neg-classical} with the two (mutually dual) situations depicted in Fig.~\ref{fig:neg-nonclassical} for non-classical negations. As illustrated, employing paraconsistent negations we can represent \textit{truth-gluts}, i.e., situations in which, for some $A$, both it and its negation $\neg A$ are the case (e.g., when we have contradictory information about $A$). Analogously, employing paracomplete negations we can represent \textit{truth-gaps}, i.e., situations in which neither $A$ nor its negation $\neg A$ is the case (e.g., when we have only partial information as to whether $A$ holds).

Thus, in our approach, paraconsistency and paracompleteness can be seen as dual notions.\footnote{Such a duality naturally follows from our consideration of lattice-based semantics for families of paraconsistent and paracomplete logics. As we will explore later, this frequently results in `dual pairs' of characterizations and other meta-theoretical results.} From a logical point of view:
\begin{align*}
\text{(paraconsistency)}\;\;\;\; A ~\land~ \neg A \;&\nvdash\; \bot \\
	\top \;&\nvdash\; A ~\lor~ \neg A \;\;\;\;\text{(paracompleteness)}.
\end{align*}
As previously mentioned, given Boolean conjunction and disjunction, it becomes necessary to relinquish either TND or ECQ (or both) for any property of negation to become `negotiable'. There are many compelling reasons to feature a `weak' negation in a logical system, and these have been thoroughly discussed in the literature on non-classical logics (see e.g.~\cite{BookCC16} and the references therein). We have listed several properties of negation in Table~\ref{table:negation-properties} that have been the subject of our investigations (see \cite{AFP}).

\begin{table}
	\centering
\begin{tabular}{ |c|c| } 
\hline			
\textbf{description} & \textbf{schema} \\ 
\hline
`weak' TND& $\neg B \vdash A \lor \neg A$  \\
`weak' ECQ& $A \land \neg A \vdash \neg B$  \\
law of non-contradiction& $\vdash \neg(A \land \neg A)$  \\
de Morgan 1& $\neg(A \lor B) \qdash \neg A \land \neg B$  \\
de Morgan 2& $\neg(A \land B) \qdash \neg A \lor \neg B$  \\ 
double negation & $A \qdash \neg\neg A$ \\ 
`weak' double negation & $\vdash A ~\Longleftrightarrow~ \vdash \neg\neg A$ \\ 
contraposition 1 & $A \limp B \qdash \neg B \limp \neg A $ \\ 
contraposition 2 & $A \limp \neg B \qdash B \limp \neg A $ \\
`weak' contraposition 1 & $A \vdash B \Longleftrightarrow \neg B \vdash \neg A $ \\ 
`weak' contraposition 2 & $A \vdash \neg B \Longleftrightarrow B \vdash \neg A $ \\ 
disjunctive syllogism & $A \lor B \qdash \neg A \limp B$  \\
\hline
\end{tabular}
\caption{Some `negotiable' properties of negation. Note that each direction ($\vdash$ resp.~$\dashv$ or $\Longrightarrow$ resp.~$\Longleftarrow$) is a different property that should be considered separately.}
\label{table:negation-properties}
\end{table}
As mentioned previously, one of our aims has been to facilitate automated reasoning with \textit{non-classical} logical formalisms.
The task of designing a custom logical system for a particular modeling problem (e.g., in AI knowledge representation) requires exploring a large space of logical inferences for different combinations of properties of the connectives in a target logic. Even for a single unary negation(-like) operation (cf.~Table~\ref{table:negation-properties}) the space  to be explored is too vast for a systematic and \textit{reliable} exploration using traditional pen-and-paper proof methods.\footnote{Some readers, being skilled logicians themselves, might feel capable of reproducing or verifying by hand all the results presented in our formalization work (see \cite{AFP}) for each combination of assumptions. The questions are rather: do they truly have the time? how reliable are those proofs and who will check them? The reader might possess dependable intuitions concerning classical, intuitionistic, or even some well-known systems of paraconsistent logic, but these intuitions won't help (in fact they might hinder) in deriving or verifying proofs in other, more exotic systems.}  In our approach, state-of-the-art automated theorem provers and (counter)model generators can be harnessed to handle most of the proof and refutation heavy lifting. This way, we can systematically explore the \textit{minimal} sets of semantic conditions that are necessary for interesting (meta)theorems to hold.\footnote{See \cite{CICM} for an exemplary application of this methodology in topology.}

\subsection{Recovery Operators}\label{subsec:prelim-recovery}

Surprisingly (or not) paraconsistent and paracomplete logics are not necessarily `weaker'\footnote{Weak(er) in the sense of validating less theorems, licensing less inferences, and so on.} than classical logic. In fact, some paraconsistent systems in the literature are strictly more expressive than classical logic: they embed the latter as a subsystem. Among such systems we have the families of paraconsistent Logics of Formal Inconsistency (LFIs) \cite{CM,BookCC16,CCR2020} and paracomplete Logics of Formal Undeterminedness (LFUs) \cite{Marcos2005,CCR2020}.
LFIs and LFUs feature additional unary connectives ($\circ$ and $\lstar$, respectively) allowing us to recover classical properties in a `sentence-wise' fashion. Following \cite{CCR2020} we refer to them as \textit{recovery operators}. Their modus operandi is described below.\footnote{We assume a fully \textit{structural} setting (in Gentzen's sense), and thus the comma-separator  `,' in a sequent behaves as a conjunction (resp.~disjunction) on the left- (resp.~right-) hand-side.}

LFIs, being paraconsistent, do not validate ECQ (aka.~`principle of explosion'), i.e.
$$A ,\neg A \nvdash \Delta \,.$$

Instead, LFIs validate the so-called `principle of gentle explosion', namely\footnote{In the literature on LFIs \cite{CM,BookCC16} an additional restriction is considered, in order for this principle to hold in a non-trivial way: there exist formulas $\alpha$, $\beta$ such that $\circ \alpha , \alpha \nvdash \beta$ and $\circ \alpha , \neg \alpha \nvdash \beta$.}
$$\circ A ,  A ,\neg A  \vdash \Delta \,.$$

The property above suggests reading the LFIs' operator $\circ$ as a kind of \textit{consistency operator}, so that the formula ${\circ}A$ reads as ``A is consistent'' \cite{CM}. Thus, under the assumption that (the evidence for) $A$ is consistent, i.e.~${\circ}A$, we have that from $A \land \neg A$ anything does follow. 

LFUs, being paracomplete, do not validate the law of excluded middle (TND), i.e.
$$ \Gamma  \nvdash A ,\neg A \,.$$

Dually to LFIs, LFUs validate instead
$$ \Gamma  \vdash {\lbstar}A ,  A ,\neg A \;\;\;\;\;\text{i.e.}\;\;\; \Gamma, {\lstar}A \vdash A ,\neg A$$
where $\lbstar A ~=~ {\cmpl}{\lstar}A$ for classical negation `$\cmpl$' (which is in fact definable inside the LFIs/LFUs considered here, as discussed below). 
Analogously, Marcos \cite{Marcos2005} proposes to read ${\lstar}A$ as ``A is determined''. Thus, assuming that (the evidence for) $A$ is determined (i.e.~complete or exhaustive), we will always have that $A \lor \neg A$ holds.

Drawing upon these intuitive readings we can think of the operators $\circ$ and $\lstar$ as sort of `quality seals' for propositions. For example, they can offer insights into the `trustworthiness' level of evidential support for propositions: $\circ A$ indicates that the evidence for $A$ is non-contradictory, while $\lstar A$ suggests that the evidence for $A$ is non-partial. We refer the reader to \cite{CR2017} for a discussion of the epistemic interpretation of LFIs and LFUs.

Enhancing a logic with recovery operators such as $\circ$ and $\lstar$ not only serves to recover the properties of ECQ and TND as shown above. In fact, other properties of negation can  be also be recovered this way. For example, in the LFI system \textbf{mbC} \cite{BookCC16} contraposition (among others) is generally not valid, e.g.
$$A \limp B \nvdash_{\textbf{mbC}} \neg B \limp \neg A \,.$$

From this we can correctly infer that the \textbf{mbC} negation ($\neg$) is indeed very `weak'. However, we shall not extend this judgment to the whole logical system. On the one hand, we can always recover classical properties in a sentence-wise fashion by employing the consistency operator $\circ$, thus for the case above we have
$$\circ B , A \limp B \vdash_{\textbf{mbC}} \neg B \limp \neg A \,.$$

On the other hand, the additional expressivity obtained via the recovery operator $\circ$ allows us to define, in systems extending \textbf{mbC}, a fully classical negation inside the logic as ${\cmpl}A = A \limp \bot$, where $\bot$ is any formula of the form $B \land \neg B \land \circ B$. In this sense, we can say that, in fact, those LFIs `extend' or `embed' classical logic. Of course, an analogous (dual) argument can be made for LFUs too.

We finish this section by noting that recent work \cite{CCF2021} has provided an algebraic semantics for extensions of the LFI system \textbf{mbC} that satisfy the property of \textit{replacement} (of provable equivalents). The minimal system in this new family has been suggestively called \textbf{RmbC}. The semantics provided in that work for \textbf{RmbC} and its extensions are based upon so-called \textit{Boolean algebras with LFI operators} (BALFIs). The algebraic semantics introduced in the present paper (based on topological Boolean algebras, cf.~\S\ref{sec:investigations}) has, in fact, grown out of an effort to generalize the neighborhood structures introduced as representations of BALFIs in \cite[\S5]{CCF2021}.

In the sequel we will always assume that our non-classical logical systems are \textit{self-extensional}, i.e., they satisfy the replacement property.

\subsection{Topological Boolean Algebras}\label{subsec:prelim-TBA}

We introduce in \S\ref{sec:investigations} a family of mathematical structures that, we argue, provide a natural semantics for some families of LFIs and LFUs, and for which we employ the umbrella term: \textit{topological Boolean algebras} (TBAs).\footnote{We have chosen this name inspired by Rasiowa \& Sikorski \cite{TMMT} who exemplarily employ structures of this kind to provide semantics for non-classical logics. However, as will be seen, we generalize its meaning to cover not only (`weakenings' of) closure and interior algebras, but also Boolean algebras featuring other operations of a somewhat topological character. The term TBAs, in our sense, is left deliberately vague. We make no attempt to define it.} We characterize TBAs rather informally as Boolean algebras extended with one or more operations satisfying some (but not necessarily all) of the properties traditionally associated with well-known topological operations on sets. As for the unary case, paradigmatic cases of such operations (operators) are the topological \textit{closure} and its dual the \textit{interior} \cite{Kuratowski1}. Moreover, other set-operations can also be interpreted as topological, in particular the \textit{border}, \textit{frontier} and \textit{exterior} \cite{Zarycki1}, as well as the \textit{derived-set} (aka.~\textit{derivative}) \cite{Kuratowski1,Zarycki3,AOT}, Hausdorff's \textit{residue} \cite{Hausdorff} and Cantor's \textit{coherence} \cite{Zarycki2}. They all count as operators in our sense of the word.

The first and most paradigmatic examples of TBAs in the literature are \textit{closure algebras}, i.e., Boolean algebras extended with a (topological) closure operator. They were in fact the structures studied in the seminal work by McKinsey \& Tarski \cite{AOT}. The corresponding closure operator and its dual (interior) were famously axiomatized by Kuratowski \cite{Kuratowski1}.

Readers familiarized with basic topology will recall the common definition of topological spaces as algebras of open (resp.~closed) sets, for which an interior (resp.~closure) operation can be defined in the usual way. However, only a few might be familiarized with the definition of topological spaces as closure algebras, as presented, e.g., by McKinsey \& Tarski \cite{AOT}. Intuitively, the equivalence of both characterizations should not come as a surprise, since, together with the simple 2-element lattice $\{\top, \bot\}$, algebras of sets (aka.~\textit{fields of sets}) provide the paradigmatic examples of Boolean algebras.\footnote{This has of course its roots in the celebrated representation theorems by Birkhoff and Stone.}

We recall below some characterizations of topological operators that will be relevant in our discussion in \S\ref{sec:investigations}. They are intended as additional unary operations on top of a Boolean algebra with carrier $\A$, signature $\langle \land, \lor, \cmpl, \top, \bot \rangle$, and ordered by $\leq$. We provide some names and mnemonics for their axiomatic conditions (\texttt{ADDI}tivity, \texttt{IDEM}potence, etc.) to facilitate references.

\begin{definition}[Closure Operator]\label{def:closure}
\normalfont
A unary operation $\C$ 
is called a (Kuratowski) \emph{closure operator} when, for arbitrary $A, B \in \A$, it is
\begin{align*}
&(\texttt{C1}) \;\;\;\text{additive} \;(\ADDI):& \C (A \lor B) &= \C(A) \lor \C(B) \\
&(\texttt{C2}) \;\;\;\text{expansive} \;(\EXPN):& A &\leq \C (A) \\
&(\texttt{C3}) \;\;\;\text{normal} \;(\NORM):& \C (\bot) &= \bot \\
&(\texttt{C4}) \;\;\;\text{idempotent} \;(\IDEM):& \C(\C(A))  &= \C(A) \,.
\end{align*}
A Boolean algebra extended with a closure operator 
is called a \emph{closure algebra}.
\end{definition}

\begin{definition}[Interior Operator]\label{def:interior}
\normalfont
A unary operation $\I$ 
is called an \emph{interior operator} when, for arbitrary $A, B \in \A$, it is
	\begin{align*}
		&(\texttt{I1}) \;\;\;\text{multiplicative} \;(\MULT):& \I (A \land B) &= \I(A) \land \I(B) \\
		&(\texttt{I2}) \;\;\;\text{contractive} \;(\CNTR):& \I(A) &\leq A \\
		&(\texttt{I3}) \;\;\;\text{dual-normal} \;(\DNRM):&  \I (\top) &= \top \\
		&(\texttt{I4}) \;\;\;\text{idempotent} \;(\IDEM):&  \I(A) &= \I(\I(A)) \,.
	\end{align*}
A Boolean algebra extended with an interior operator 
is called an \emph{interior algebra}.
\end{definition}

Closure and interior operators are said to be \textit{dual}, since given one of them, say $\C$, we can define the other, say $\I(A) \equdef \cmpl\C(\cmpl A)$ for any $A$. Let us make this relationship official by defining the dual $\supd{(\cdot)}$ of an operator $\vfi$ as $\supd{\vfi}(A) \equdef \cmpl\vfi(\cmpl A)$.

It is well known that closure/interior algebras and topological spaces are two sides of the same coin. This insight has been leveraged, at least since the seminal work by McKinsey \& Tarski in the 1940's \cite{AOT}, to provide topological semantics for intuitionistic and modal logics (see \cite{Esakia} for a brief survey).
In fact, by employing different subsets of the closure conditions listed above (\texttt{C1-4}) we can also characterize some interesting operators, to name just a few:
\begin{itemize}
	\item The conditions \texttt{C1} and \texttt{C3} (dually: \texttt{I1} and \texttt{I3}) suffice to characterize a modal \textit{possibility} (dually: \textit{necessity}) operator.\footnote{Cf.~J\'onsson \& Tarski's \textit{Boolean Algebras with Operators} \cite{BAO}. This seminal work provided, in today's terminology, an algebraic semantics for (extensions of) normal modal logic \textbf{K}. Observe that condition \texttt{I2} (\texttt{C2}) and the left-to-right direction ($\leq$) of \texttt{I4} (\texttt{C4}) correspond to the (duals of) modal axioms $T$ and $4$ respectively. Hence taking all Kuratowski conditions together corresponds to axiomatizing modal logic \textbf{S4}.}
	\item The conditions \texttt{C1}, \texttt{C2} and \texttt{C3} (leaving out idempotence: \texttt{C4}) characterize a preclosure (aka.~\v{C}ech closure) operator.
	\item The ubiquitous Moore (aka.~hull) closure is characterized by the right-to-left direction ($\geq$) in \texttt{C1} (equivalent to monotonicity), together with \texttt{C2} and \texttt{C4}.
	\item  Another topological operator, the derived-set (aka.~derivative) operation \cite{Kuratowski1,Zarycki3,AOT} can be characterized by 
	\texttt{C1}, \texttt{C3}, and the left-to-right direction ($\leq$) in \texttt{C4}.
\end{itemize}

In instances like those above where we utilize only a subset of the Kuratowski closure conditions (\texttt{C1-4}) to constrain a given unary operation, say $\C$, we might rigorously refer to it, for example, as a `generalized closure(-like) operator'. Analogous designations could be devised for unary operations satisfying only a subset of the interior conditions (\texttt{I1-4}) or the other sets of conditions to be introduced later. To circumvent such contrived terminologies, we will adopt the terms `closure', `interior', etc., more loosely, so they also apply to `weaker' operators. We will let the context determine how operators are referred to, as long as it does not lead to confusion.

Bridging between the notions of closure and interior we find the notion of \textit{exterior} (of a set), which has been axiomatized by Zarycki \cite{Zarycki1} as follows.

\begin{definition}[Exterior Operator]\label{def:exterior}
\normalfont
A unary operation $\E$ 
is called an \emph{exterior operator} when, for arbitrary $A, B \in \A$, it satisfies
\begin{align*}
	&(\texttt{E1}) \;\;\;& \E(A \lor B) &= \E(A) \land \E(B) \\
	&(\texttt{E2}) \;\;\;& \E(A) &\leq \cmpl A  \\
	&(\texttt{E3}) \;\;\;& \E(\bot) &= \top \\
	&(\texttt{E4}) \;\;\;& \E(\cmpl\E(A))  &= \E(A) \,.
\end{align*}
We may call a Boolean algebra extended with an exterior operator 
an \emph{exterior algebra}.
\end{definition}

Note that the axiomatic conditions for the exterior operator display a certain antagonism towards those for interior and closure (e.g., \texttt{E1} will be referred to as `anti-additivity' later in \S\ref{sec:investigations}). From another perspective, these concepts are interrelated since, given a closure $\C$ (resp.~an interior $\I$), an exterior operator $\E$ can be defined as $\E(A) \equdef \cmpl\C(A)$ (resp.~$\E(A) \equdef \I(\cmpl A)$). Therefore, as with interior operators, exterior operators can be readily obtained by composing closure operators with the complement. In fact, Kuratowski \cite{Kuratowski1} famously demonstrated that the number of operators achievable in this manner is limited to 14.\footnote{Of course, this result hinges on assuming (all?) Kuratowski closure conditions.}

Other topological operators axiomatized by Zarycki in \cite{Zarycki1} were the \textit{border} and the \textit{frontier} (aka.~\textit{boundary}).\footnote{They have been called `bord' and `fronti\`ere', respectively, in the early writings (in French) of Zarycki, Kuratowski, and other mathematicians of the era. The term `boundary' is somewhat more modern and widespread, but we abstain from using it here, as it is commonly employed for another related concept.}   The former, in particular, will play an important role in our investigations on recovery operators in \S\ref{subsec:recovery}.

\begin{definition}[Border Operator] \label{def:border}
\normalfont
A unary operation $\B$ 
is called a \emph{border operator} when, for arbitrary $A, B \in \A$, it satisfies\footnote{We note that the axiomatization originally introduced by Zarycki \cite{Zarycki1} features only three conditions, since it collapses our \texttt{B1} and \texttt{B2} into one equivalent formula: $\B(A \land B) = (A \land \B(B)) \lor (B \land \B(A))$. Our formulation here is not only more explicit but also makes it easier to interrelate conditions for different operators with a finer granularity.}
	\begin{align*}
		&(\texttt{B1}) \;\;\;& A \land B \land \B(A \land B) &= A \land B \land (\B(A) \lor \B(B)) \\
		&(\texttt{B2}) \;\;\;& \B(A) &\leq A \\
		&(\texttt{B3}) \;\;\;& \B(\top) &= \bot \\
		&(\texttt{B4}) \;\;\;& \B(\cmpl\B(\cmpl A)) &\leq A \,.
	\end{align*}
We may call a Boolean algebra extended with a border operator 
a \emph{border algebra}.
\end{definition}

\begin{definition}[Frontier Operator] \label{def:frontier}
\normalfont
A unary operation $\F$ 
is called a \emph{frontier operator} when, for arbitrary $A, B \in \A$, it satisfies
	\begin{align*}
		&(\texttt{F1}) \;\;\;& A \land B \land \F(A \land B) &= A \land B \land (\F(A) \lor \F(B)) \\
		&(\texttt{F2}) \;\;\;& \F(\cmpl A) &= \F(A) \\
		&(\texttt{F3}) \;\;\;& \F(\bot) &= \bot \\
		&(\texttt{F4}) \;\;\;& \F(\F(A)) &\leq \F(A) \,.
	\end{align*}
We may call a Boolean algebra extended with a frontier operator 
a \emph{frontier algebra}.
\end{definition}

From the conditions above we can discern that the border ($\B$) and frontier ($\F$) operators are related in some sense, albeit they are still quite distinct.
For instance, conditions \texttt{B1} and \texttt{F1} are identical. This condition has been dubbed in \S\ref{subsec:conditions} (somewhat awkwardly) `anti-multiplicativity relative to $(A \land B)$'. However, condition \texttt{B2}, which is identical to \texttt{I2} (contractiveness), differs from condition \texttt{F2} (symmetry wrt.~complement). Condition \texttt{F3} is identical to \texttt{C3} (normality), while \texttt{B3} has a reminiscent form (it has been dubbed, awkwardly again, `anti-dual-normality' in \S\ref{subsec:conditions}). Finally, \texttt{F4} corresponds to the $\leq$ direction of \texttt{C4} (idempotence), while \texttt{B4} has a quite different form.\footnote{Border operators are in fact idempotent, i.e., $\B(\B(A)) = \B(A)$; this follows already from \texttt{B1} and \texttt{B2}. Furthermore, \texttt{B4} is related to a variant of what we refer to as `anti-idempotency' in \S\ref{subsec:conditions}.} These sets of axiomatic conditions and their interrelationships will be systematically explored in \S\ref{subsec:conditions}.

We have quoted above the famous result by Kuratowski \cite{Kuratowski1}, which states that a maximum of 14 different operators can be defined by composing the complement ($\cmpl$) with a (fully axiomatized) closure operator $\C$ (and, by extension, with $\I$ and $\E$). In the same spirit, Zarycki \cite{Zarycki1} demonstrates that any operator obtained via composition of the complement with the frontier operator ($\F$) must coincide with one of the following six operations, which given $A$ return respectively: $A, \F(A), \F(\F(A)), \cmpl A, \cmpl\F(A), \cmpl\F(\F(A))$. In contrast, Zarycki \cite{Zarycki1} also indicates that an infinite number of operators can be obtained in this manner when considering the border operator ($\B$).\footnote{This is because there exist sets with a transfinite family of residues, where the residue of a set $A$ corresponds to $\B(\cmpl\B(\cmpl A))$; cf.~Hausdorff \cite{Hausdorff}.}

\begin{figure}
	\centering
	\includegraphics[width=.50\textwidth]{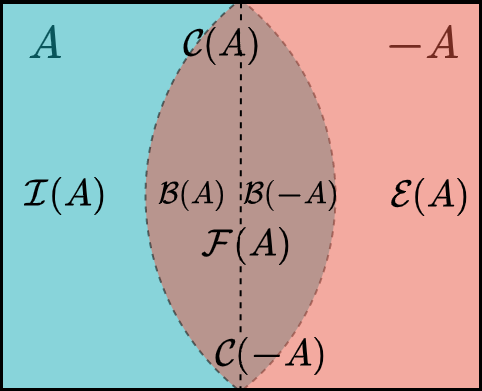}		
	\caption{Suggestive pictorial representation of topological operators and their interrelations. The blue, resp.~red, areas correspond to $\C(A)$, resp.~$\C(\bcmpl A)$; they overlap in the brownish area corresponding to $\F(A)$ ($= \F(\cmpl A)$), itself partitioned into $\B(A)$ and $\B(\cmpl A)$ (corresponding in a sense to the \textit{truth-gluts} and \textit{truth-gaps} in Fig.~\ref{fig:neg-nonclassical}). Observe also that $\E(A)$ corresponds to $\I(\cmpl A)$, and thus to $\cmpl\C(A)$.}
	\label{fig:topo-operators}
\end{figure}

Zarycki \cite{Zarycki1} also shows that closure, interior, exterior, border and frontier operators are all inter-definable, i.e., having introduced one of them as primitive we can employ it to define the others as follows (see Table \ref{table:op-interrelations} and also Fig.~\ref{fig:topo-operators} for a suggestive illustration):
\begin{table}
\centering
		\begin{tabular}{ |c|c|c|c|c|c| } 
			\hline
			\small{define\textbackslash using} & $\C$ & $\I$ & $\E$ & $\B$ & $\F$\\ 
			\hline
			$\C(A)$ & -- & $\supd{\I}(A)$ & $\cmpl\E(A)$ & $A \lor \B({\cmpl}A)$ & $A \lor \F(A)$ \\ 	
			\hline
			$\I(A)$ & $\supd{\C}(A)$ & -- & $\E(\cmpl A)$ & $A \land {\bcmpl}\B(A)$ & $A \land {\cmpl}\F(A)$ \\ 	
			\hline	
			$\E(A)$ & ${\cmpl}\C(A)$ & $\I({\cmpl}A)$ & -- & ${\cmpl}A \land \supd{\B}(A)$ & ${\cmpl}A \land {\cmpl}\F(A)$\\ 	
			\hline		
			$\B(A)$ & $A \land \C({\cmpl}A)$ & $A \land {\cmpl}\I(A)$ & $A \land \supd{\E}(A)$ & -- & $A \land \F(A)$ \\ 	
			\hline
			$\F(A)$ & $\C(A) \land \C({\cmpl}A)$ & ${\cmpl}(\I(A) \lor \I(\cmpl A))$ & ${\cmpl}(\E(A) \lor \E(\cmpl A))$ & $\B(A) \lor \B({\cmpl}A)$ & -- \\ 
			\hline
		\end{tabular}
	\caption{Inter-definitions between topological operators.}
	\label{table:op-interrelations}
\end{table}

It is worth noting that inter-definitions involving \textit{only} $\C$, $\I$ and $\E$ obtain already without assuming any restriction. By contrast, inter-definitions involving any of $\B$ or $\F$ always assume the second Kuratowski condition (or its respective counterpart: \texttt{C2}, \texttt{I2}, \texttt{E2}, \texttt{B2} or \texttt{F2}). For example, recalling Table \ref{table:op-interrelations}, converting an operator, say $\B$, into an interior $\I_\B$ (entry $\langle 4,2 \rangle$) and then back to $\B$ (entry $\langle 2,4 \rangle$) requires assuming \texttt{B2}. 

As we have seen, after extending a Boolean algebra with any of the topological operators discussed above ($\C$, $\I$, $\E$, $\B$, $\F$), we obtain a TBA where all other operators are definable. Moreover, as will be shown in \S\ref{sec:investigations}, several other operators can be defined by employing, among others, function composition, for which we employ the notation $(\vfi\,\circ\,\psi)(A) \equdef \vfi(\psi(A))$ and write $\vfi^{n+1}$ for $\vfi$ composed with itself $n$-times.

Furthermore, Zarycki has also shown in \cite{Zarycki1} how to inter-derive all corresponding axiom systems for the above operators on top of algebras of sets. We can axiomatize, e.g., an operator $\F$ with the frontier conditions \texttt{F1-4} (thus obtaining a frontier algebra) and derive, as theorems, the closure conditions \texttt{C1-4} for a closure operator defined as $\C(A) \equdef A \lor \F(A)$.
We have employed in \cite{AFP} automated tools to elicit \textit{minimal} sets of conditions under which these correspondences hold. We will revisit this issue in \S\ref{subsec:conditions}.

\textit{Fixed points} of topological operators are also interesting to consider. We employ below the notation $\texttt{fp}(\vfi,A) \equdef \vfi(A) = A$ (read: ``$A$ is a fixed point of $\vfi$''), recalling that fixed points of $\C$, resp.~$\I$, are called \textit{closed}, resp.~\textit{open}, in the literature. In a similar vein, we can consider the fixed points of the operators discussed above and observe that they satisfy interesting topological properties (with operators inter-defined as in Table~\ref{table:op-interrelations}).
\begin{itemize}
	\item $A$ is \textit{open} iff its border is the least element (`bottom'): $\texttt{fp}(\I,A)$ iff $\B(A) = \bot$.
	\item $A$ is \textit{closed} iff the border of its complement is `bottom': $\texttt{fp}(\C,A)$ iff $\B(\cmpl A) = \bot$.
	\item $A$ is \textit{clopen} (both closed and open) iff its frontier is `bottom': $\texttt{fp}(\C,A)$ and $\texttt{fp}(\I,A)$ iff $\F(A) = \bot$.
	\item $\E$ has no fixed points. However, fixed points of $\E^2$ correspond to \textit{regular open} elements: $\texttt{fp}(\E^2,A)$ iff $\texttt{fp}(\I{\circ}\C,A)$; dually, fixed points of $(\E^d)^2$ correspond to \textit{regular closed} elements.
	\item Fixed points of $\B$ correspond to \textit{boundary} elements (being analogous to the so-called `boundary sets' in point-set topology): $\texttt{fp}(\B,A)$ iff $\I(A) = \bot$.
	\item Fixed points of $\B^d$ correspond to \textit{dense} elements: $\texttt{fp}(\B^d,A)$ iff $\C(A) = \top$.
	\item Fixed points of $\F$ correspond to those elements that are \textit{closed} and \textit{nowhere-dense}: $\texttt{fp}(\F,A)$ iff $\texttt{fp}(\C,A)$ and ${\I(\C(A)) = \bot}$. Conversely, \textit{nowhere-dense} elements can be characterized as those whose closure is a fixed point of $\F$ or, alternatively, as the elements that are smaller than the frontier of their closure: ${\I(\C(A)) = \bot}$ \,iff\, $\texttt{fp}(\F,\C(A))$ \,iff\, $A \leq \F(\C(A))$.

\end{itemize}

Certainly, in the topological literature, it is always presupposed that operators satisfy all of the corresponding axiomatic (e.g., Kuratowski) conditions. A distinctive feature of our approach is that these axiomatic conditions become `negotiable',\footnote{As previously mentioned, this is even to the extent that qualifications like `closure' or `closed' might become challenging to justify. We stick to them, however.} allowing us to pose interesting `reverse-mathematical' sort of questions involving the \textit{minimal} set of conditions required, e.g., for the relationships listed above to hold. Some of these questions have been addressed in a topological context in previous work \cite{CICM}. We argue that, in order to answer such questions \textit{at scale}, we need to leverage computational tools like automated theorem provers and model generators. For instance, we employ the automated tools integrated into the Isabelle/HOL proof assistant. Automated theorem provers are utilized (via \textit{Sledgehammer} \cite{blanchette2016hammering}) to verify entailment and (counter)model generators (like \textit{Nitpick} \cite{Nitpick}) are employed to verify satisfiability and to find counter-examples to non-theorems (e.g., when some necessary assumption is lacking).

\subsection{Degree-preserving Logical Consequence}\label{subsec:prelim-logical-consequence}

Finally, having introduced our target classes of algebraic structures (TBAs), we discuss how they can serve to provide a semantics for families of non-classical logics such as LFIs and LFUs. We note that our target logics are assumed to be structural (in Gentzen's sense) and Tarskian. They also satisfy the property of replacement (of provable equivalents), i.e., they are self-extensional.

We begin with the classical connectives. As expected, conjunctions and disjunctions are interpreted by meets and joins ($\land$, $\lor$), respectively. We may also conveniently include $0$-ary `top' and `bottom' connectives in our logic, interpreted by their algebraic counterparts ($\top$, $\bot$). Classical negation may or may not be part of our language; if it is, then it becomes interpreted by the complement operation ($\cmpl$). Implication is interpreted by the corresponding algebraic connective (or defined as shorthand, e.g., using classical negation and disjunction).
We can optionally introduce `modalities' in our language, which are to be interpreted by operators (e.g., $\Diamond$ interpreted as $\C$ and \,$\Box$\, as $\I$).

A distinguishing feature of our approach is that non-classical negations (paraconsistent or paracomplete) and their corresponding recovery operators are interpreted by operators that are suitably interrelated. 
This will be discussed in \S\ref{subsec:negations}-\ref{subsec:recovery}.

Importantly, we interpret logical entailment using the algebra's ordering relation. Thus, e.g.,
$$A ,  B \vdash C \,(,  D )$$
becomes interpreted as\footnote{Recall that, since we assume a fully \textit{structural} setting (in Gentzen's sense), the comma-separator  `,' in a sequent behaves as a conjunction (disjunction) on the left- (right-) hand-side. Families of non-classical logics for which this assumption does not hold are called \textit{substructural logics} in the literature. Their treatment lies beyond the scope of the present paper.}  
$$A \land B \leq  C \,(\lor\, D ) \,.$$

This approach towards characterizing logical consequence for (semi-)lattice-based algebraic semantics is  well-known in the literature (e.g., in fuzzy logics), among others, under the name of \textit{degree-preserving} consequence \cite{bou2009logics}, and corresponds, from an algebraic point of view, to the standard notion of consequence employed in modal logic, so-called \textit{local} consequence.
This can be contrasted with \textit{truth-preserving} (aka.~\textit{global}) consequence, in which, e.g.,
$$A ,  B \vdash_{\texttt{g}}  C \,(,  D )  $$
becomes interpreted as  
$$ A = \top  \text{\;\;and\;\;}  B = \top \text{\;\;implies\;\;}  C = \top \,\,(\text{or\,\,} D = \top) \,.$$
Clearly, theorems coincide in both approaches, where 
$$\vdash A \text{\;\;\;iff\;\;\;} A = \top \text{\;\;\;iff\;\;\;} \vdash_g A \,.$$

However, degree-preserving consequence is stronger than truth-preserving consequence. It is easy to see that $A \vdash B$ implies $A \vdash_g B$ but not vice versa.

A characteristic of the notion of degree-preserving consequence is that any logic featuring an intuitionistic (or classical) implication automatically satisfies the deduction theorem (wrt.~that connective). Hence we have that
$$A \vdash B \text{\;\;\;iff\;\;\;} A \leq B \text{\;\;\;iff\;\;\;} A \shortrightarrow B = \top \text{\;\;\;iff\;\;\;} \vdash A \shortrightarrow B \,.$$

\section{Semantical Investigations}\label{sec:investigations}

The main results of our work are presented in this section, which is structured as follows:
We begin in \S\ref{subsec:SSE} by discussing the \textit{shallow semantical embedding} approach, which we employ to develop a theory of unary operations (operators) on top of Boolean algebras, encoded in \S\ref{subsec:BA} as algebras of sets. In \S\ref{subsec:operators} we introduce some simple transformations between operators giving rise to an interesting commutative diagram in the form of an abstract `cube of opposition'. We show in \S\ref{subsec:conditions} how different sets of axiomatic conditions (\`a la Kuratowski) become `translated' as we transform operators into each other (moving along the edges of the cube). In \S\ref{subsec:cube-concrete} we discuss some concrete operators, as featured in topological Boolean algebras (TBAs), and use them to instantiate a \textit{topological cube of opposition}. We then show how non-classical (paraconsistent and paracomplete) negations (in \S\ref{subsec:negations}) and recovery operators (in \S\ref{subsec:recovery}) can be defined in terms of topological operations and their transformations. Finally, we provide in \S\ref{subsec:quantifiers} a uniform characterization of propositional, first-order, and higher-order quantification (restricted also to constant and varying domains) and discuss the Barcan formula (and its converse).

\subsection{Higher-order Logic (HOL) as a Meta-Logic}\label{subsec:SSE}

We have mentioned previously that our target non-classical logics are to be introduced as a family of \textit{object logics} by encoding logical connectives as terms in a \textit{classical higher-order meta-logic}, in which their truth conditions are stated and semantic conditions axiomatized. This formal meta-logic is suggestively referred to as HOL in this article.\footnote{HOL is in fact an acronym for (classical) higher-order logic. It is traditionally employed to refer to a family of extensions of Church's simple type theory \cite{SEPTT} commonly employed in interactive theorem proving, starting with the eponymous \textit{HOL} proof assistant \cite{gordon1988hol} and continuing with descendant systems like \textit{HOL-Light} \cite{harrison2009hol} and \textit{Isabelle/HOL} \cite{Isabelle}.}

HOL is a \textit{logic of functions}. As a higher-order logic, HOL is a very expressive logical system that allows for quantification over predicate and function variables. In the context of automated reasoning, the term \textit{higher-order logic} (in general) and HOL \cite{gordon1985hol} (in particular) customarily refers to \textit{classical} logical systems extending the simply typed $\lambda$-calculus, as originally introduced by Church \cite{Church40}, and often referred to as \textit{simple type theory} \cite{SEPTT}.

To keep this article sufficiently self-contained we will succinctly introduce HOL's syntax and briefly motivate its semantics. We refer the interested reader to the literature on HOL semantics \cite{henkin1950completeness,andrewsBook,J6}, proof systems and their automation \cite{B5}.

As a typed logic, all terms of HOL get assigned a fixed and unique \textit{type}, commonly written as a subscript (i.e., the term $t_\alpha$ has $\alpha$ as its type). The use of types in HOL aims at eliminating the paradoxes and inconsistencies commonly found in untyped systems.
The set $\T$ of HOL's types is inductively generated from a set of base types $\B\T$ and the function type constructor $\ar$ (written as a right-associative infix operator\footnote{We can e.g.~omit parentheses in the type $\alpha \ar (\beta \ar \gamma)$ but not in the type $(\alpha \ar \beta) \ar \gamma$.}, see Fig.~\ref{fig:HOL-grammar}).
Traditionally, the generating set $\B\T$ is taken to include at least two base types, $\B\T \supseteq \{\iota, \bool\}$, where $\iota$ is intuitively interpreted as the type of \textit{individuals} and $\bool$ as the type of Boolean \textit{truth-values}. 
For instance, $\bool$,  $\bool \ar \bool$, $\iota\,{\ar}\,\iota$ and $\iota\,{\ar}\,\iota\,{\ar}\,\bool$ are types. Further base types may be added to $\B\T$ as needed.

\vspace*{-20pt}
\begin{figure}
	\normalsize	\centering
\begin{align*}
\alpha,\, \beta \;&::=\;\; \tau \in \B\T  \;\;\; | \;\;\; \alpha \,\ar\, \beta \\
s,\, t \,\;&::=\;\; c_\alpha \in \Const \;\; | \;\; x_\alpha \in \Var \;\; | \;\; \left(\lambda x_\alpha.\, s_\beta\right)_{\alpha\ar\beta}
\;\; | \;\; \left(s_{\alpha\ar\beta} \; t_\alpha\right)_\beta
\end{align*}
\vspace*{-15pt}
\caption{Formal grammar for HOL types ($\alpha, \beta \in \T$) and terms ($s, t \in \Lang$).}
\label{fig:HOL-grammar}
\end{figure}
\vspace*{-10pt}
In contrast to traditional first-order logic, HOL is a logic of terms only (terms of type $\bool$ are customarily referred to as `formulas').
The set $\Lang$ of HOL's \emph{terms} is inductively defined
starting from a collection of typed constant symbols ($\Const$) and typed
variable symbols ($\Var$) using the constructors \emph{function abstraction}
and \emph{function application},
thereby obeying type constraints as indicated (see Fig.~\ref{fig:HOL-grammar}).
Type subscripts and parentheses are usually omitted to improve readability, if obvious from the context or irrelevant.
Moreover, we shall assume that $\Const$ contains the symbols $\Q^\alpha_{\alpha\ar\alpha\ar \bool}$ (infix notation: $=^\alpha$) for each $\alpha \in \T$, which are interpreted as (primitive) \textit{equality}. Logical symbols, including
conjunction ($\land_{\bool\ar\bool\ar\bool}$), disjunction ($\lor_{\bool\ar\bool\ar\bool}$), 
material implication ($\longrightarrow_{\bool\ar\bool\ar\bool}$),
negation ($\neg_{\bool\ar\bool}$), Boolean constants ($\texttt{true}_\bool$\,,\,$\texttt{false}_\bool$)  and universal quantification for predicates
over type $\alpha$ ($\Pi^\alpha_{(\alpha\ar\bool)\ar\bool}$) can be introduced as
primitive constant symbols in $\Const$ (and thus given fixed intended interpretations), or, alternatively, they are often defined as abbreviations in terms of equality $\Q^\alpha$ \cite{SEPTT,andrewsBook}. In any case, we refer to them as HOL's \emph{logical connectives}.
For convenience, \emph{binder notation} is introduced for quantifiers, such that $\forall^\alpha x_{\alpha}.\,s_\bool$ can be used as an abbreviation for
$\Pi^\alpha_{(\alpha \ar \bool)\ar \bool}\,\lambda x_{\alpha}.\,s_{\bool}$  for each $\alpha \in \T$.
Additionally, \emph{description} and  \textit{choice} operators may be added to the language \cite{SEPTT,andrewsBook}.

Thus, HOL provides $\lambda$-notation as an expressive variable-binding mechanism to represent unnamed functions (\textit{function abstraction}), which can also be employed to encode predicates and sets via their \textit{characteristic functions}. Note that, by construction, HOL syntax only admits functions that take one parameter; e.g., the function that adds $2$ to a given number can be represented as $(\lambda y.\, 2 + y)$. Functions having two or more arguments must be encoded in HOL in terms of one-argument functions. In this case the values returned when applying these functions are themselves functions, which are subsequently applied to the next argument; e.g., the function that adds two numbers can be encoded as $(\lambda x.\, (\lambda y.\, x + y))$. This technique, introduced in \cite{schoenfinkel1924}, is commonly called \textit{currying}.
Hence, HOL can represent \emph{partial application} for n-ary functions; e.g., the term $(\lambda y.\, 2 + y)$ can be used to represent the partial application of the binary function `+' to the argument `2'.

For our current purposes, we shall omit the detailed presentation of HOL semantics and of its numerous proof systems, and instead refer the interested reader to the literature (e.g., \cite{J6,B5,SEPTT,andrewsBook}). We restrict ourselves to mentioning that in set-based approaches to HOL's semantics each model is based upon a \textit{frame}: a collection $\{\D_\alpha\}_{\alpha \in \T}$ of non-empty sets, called \textit{domains} (for each type $\alpha$). Since HOL is classical, $\D_\bool$ is restricted to a two-element set, say $\{T,F\}$, whereas $\D_\iota$ may have arbitrarily many elements (aka.~individuals). Expectedly, the set $\D_{\alpha \ar \beta}$ consists of functions with domain $\D_\alpha$ and codomain $\D_\beta$. In so-called \textit{standard models} \cite{SEPTT}, these domain sets are assumed to be \textit{full} (i.e.~they contain \textit{all} functions from $\D_\alpha$ to $\D_\beta$).
Broadly speaking, the main idea is to define a so-called \textit{denotation function} ($|\cdot|$) that interprets each term $s_\alpha$ as an element $|s_\alpha|$ of $\D_\alpha$ (its \textit{denotation}). As expected, a \textit{denotation function} worthy of its name must respect the intended semantics of HOL as a \textit{logic of functions}. Thus, $| s_{\alpha\ar\beta} \; t_\alpha | \in \D_\beta$ denotes the value of the function $|s_{\alpha\ar\beta}| \in \D_{\alpha\ar\beta}$ when applied to $|t_\alpha| \in \D_\alpha$, and $|\lambda x_\alpha.\, s_\beta| \in \D_{\alpha\ar\beta}$ denotes an (unique) `appropriate' function from $\D_\alpha$ to  $\D_\beta$ (see e.g.~\cite[\S2]{SEPTT} for details on how to define this). It shall be noted that the (primitive) equality constants $\Q^\alpha_{\alpha\ar\alpha\ar \bool}$ (infix: $=^\alpha$) must be given a fixed intended interpretation in such a way that $|A_\alpha =^\alpha B_\alpha|$ denotes $T \in \D_\bool$ iff $|A_\alpha|$ is identical to $|B_\alpha| \in  \D_\alpha$.
The constants $\texttt{true}_\bool$, $\texttt{false}_\bool$ are interpreted by $T, F \in \D_\bool$ respectively.
The quantifier constants $\Pi^\alpha$ are interpreted as 2nd-order predicates, i.e.~$|\Pi^\alpha| \in \D_{(\alpha \ar \bool)\ar \bool}$, that take a predicate as an argument and return $T \in \D_\bool$ iff the predicate holds for \textit{all} the objects in the domain $\D_\alpha$. It is worth noting that we assume \textit{functional extensionality} in HOL \cite{SEPTT,J6}. Therefore, quantifiers can be defined as abbreviations using equality: ${\Pi^\alpha \equdef \lambda P_{\alpha\ar\bool}.~(P =^{\alpha\ar\bool} \lambda x_\alpha.~ \texttt{true}_\bool)}$.

Logic folklore has it that, as a consequence of G\"odel's incompleteness theorems, HOL with 'standard' semantics is necessarily incomplete. By contrast, automated reasoning in HOL is usually considered with respect to so-called \textit{general models} (introduced by Henkin \cite{henkin1950completeness}) in which a meaningful notion of completeness can be achieved (see \cite{Andreka2014} for a discussion). Broadly speaking, in general models the domain sets $\D_{\alpha \ar \beta}$ are not necessarily \textit{full}, but still contain enough elements (functions) to guarantee that any term $s_{\alpha \ar \beta}$ has a denotation. Note that standard models are subsumed under general models, and thus HOL formulas proven valid with respect to the general semantics are also valid in the standard sense. Conversely, any \textit{finite} (counter)model found by a model generator is necessarily standard \cite[\S 54]{andrewsBook}.\footnote{Current model generators can only find finite (counter)models. Moreover, we shall observe that (classes of only) non-standard models cannot be characterized via HOL formulas \cite[\S55]{andrewsBook}. As a consequence, it is not clear how (if at all) the results delivered by HOL automated reasoners might ever differ from the `standard' ones. In the words of Andrews \cite[p.\,255]{andrewsBook}: \textit{``one who speaks the language of [HOL] cannot tell whether he lives in a standard or nonstandard world, even if he can answer all the questions he can ask.''}}

In the context of modal and non-classical logics, a special technique, termed shallow semantical embedding \cite{J41,J23}, has been developed to encode (quantified) non-classical logics into HOL as a meta-language, in such a way that object-logical formulas correspond to HOL terms; this is realized by directly encoding in HOL the truth conditions (semantics) for object-logical connectives as syntactic abbreviations (definitions) in basically the same way as they stand in the textbook%
\footnote{Figuratively speaking, traditional logic texts can be seen as `embedding' the semantics of object logics into natural language (e.g., English plus mathematical shorthand) as a meta-language.}
This way we can carry out semi-automated\footnote{We employ the expression `semi-automated' because the process of finding the proofs cannot be fully automated. Proof-checking in HOL is, on the other hand, a much easier problem for which very effective implementations exist. This has motivated the field of \textit{interactive theorem proving} (ITP) which aims at effectively harnessing human ingenuity for proof search, by enabling automated real-time proof-checking feedback via a graphical user interface. Proof assistants (Isabelle, Coq, Lean, etc.) are paradigmatic ITP systems.} object-logical reasoning by translation into a formal meta-logic, HOL, for which good automation support exists. 
As argued, e.g., in \cite{J41}, such a shallow embedding allows us to reuse state-of-the-art automated theorem provers and model generators for reasoning with (and about) many different sorts of non-classical logical systems in a very efficient way (avoiding, e.g., inductive definitions and proofs).

The idea of employing HOL as meta-logic to encode \textit{quantified} non-classical logics has been exemplarily discussed by Benzm\"uller \& Paulson in \cite{J23} for the case of normal multi-modal logics featuring first-order and propositional quantifiers. Their approach draws upon the `propositions as sets of worlds' paradigm from modal logic, by adding the twist of encoding sets as their characteristic functions, i.e., as total functions with a (2-valued) Boolean codomain, in such a way that a set $S$ becomes encoded as the function $S(\cdot)$ such that $a \in S$ iff $(S~a) = \texttt{true}$.
In this way, object-logical propositional connectives become easily encoded via the \textit{standard translation} of modal logic into first-order logic. On top of this, Benzm\"uller \& Paulson \cite{J23} have shown how to encode object-logical quantifiers (first-order and propositional) by lifting the meta-logical ones. The approach presented in \S\ref{subsec:quantifiers} builds upon theirs.

\subsection{Boolean Algebras}\label{subsec:BA}

A distinctive feature of our approach consists in encoding Boolean algebras and their extensions by means of their (Stone) representation as algebras of sets (aka.~`fields of sets’).
In a nutshell, this means that each element of (the carrier of) the algebra will be a set of what we refer to as \textit{points}.
Inspired by the `propositions as sets of worlds' paradigm from modal logic, we may think
of \textit{points} as being `worlds', and thus of the elements of our Boolean algebras as `propositions'. Of course, this is just one among many possible interpretations, and nothing stops us from thinking of points as any other kind of object (e.g., they can be sets, functions, sets of functions, etc.). We make use of parameters (Greek letters like $\ww$, $\itype$, $\beta$) in our types to this end.\footnote{Type parameters (e.g., $\ww$, $\itype$) are introduced also in our formalization sources \cite{AFP}, using apostrophe-leading symbols (e.g., $\text{`}w$ resp.~$\text{`}a$) which is the notation for type variables in Isabelle/HOL \cite{Isabelle}. This use of type parameters is possible thanks to Isabelle/HOL's support of \textit{rank-1} type polymorphism.}

When given a term of a $\ww$-parameterized type, we take $\ww$ as a type whose semantic domain corresponds to the domain or universe of points. For instance, the $\ww$-parametric type $\ww\ar \bool$ can be interpreted as the type of (characteristic functions of) sets of points. Given the fundamental role sets (of points) play in this work, we introduce a convenient abbreviation, $\wsig$, as (postfix) shorthand notation for the $\ww$-parametric type $\ww\ar\bool$. Often, we will simply write $\sigma$ when the actual type parameter is irrelevant or easily inferred from the context.

We will follow the convention of employing \textbf{boldface} for the encoded object-logical algebraic operations ($\band, \bimp, \bcmpl, \bbot$, etc.), ordering ($\bprec$), and equality ($\bapprox$).
We begin by encoding the latter two as binary predicates on sets, i.e., having type: $\sigma\ar\sigma\ar\bool$.
\begin{align*}
A \bprec B &\equdef\; \forall w .~(A~w) \imp (B~w) \\
A \bapprox B &\equdef\; \forall w .~(A~w) \longleftrightarrow (B~w) \;\;\;(\equ\; A \bprec B  ~\land~  B \bprec A)\\
i.e.~\;A \bapprox B &\equ\; (A ~=~ B)
\end{align*}
 It is worth noting that algebraic equality ($\bapprox$) is identical to its meta-logical counterpart ($=_{\sigma\ar\sigma\ar\bool}$) as a consequence of HOL's \textit{functional extensionality} \cite{SEPTT}.
 Observe that the binary operations meet ($\band$) and join ($\bor$) become encoded as terms of type $\sigma\ar\sigma\ar\sigma$, by reusing HOL's conjunction and disjunction, what makes them tantamount to set intersection and union, respectively. Similarly the top ($\btop$) and bottom ($\bbot$) elements are encoded as zero-ary connectives (of type $\sigma$) by  reusing the meta-logical \texttt{true} and \texttt{false} terms.
\begin{align*}
A \band B \equdef&~ \lambda w .~(A~w) \land (B~w) \;\;\; &
\btop \equdef&~ \lambda w .~\texttt{true} \\
A \bor B \equdef&~ \lambda w .~(A~w) \lor (B~w) \;\;\; &
\bbot \equdef&~ \lambda w .~\texttt{false}	
\end{align*}
Drawing upon the classicality of our meta-logic (HOL) we can define an implication connective ($\bimp$) analogously by reusing HOL's classical counterpart. Moreover, the object-logical Boolean complement ($\bcmpl$) can be equivalently encoded by reusing HOL's negation.
\begin{align*}
A \bimp B \equdef\; \lambda w .~(A~w) \imp (B~w) \;\;\;\;\;\;\;\;\;\;
\bcmpl A \equdef\; \lambda w .~\neg(A~w)  
\end{align*}
Other related operations, such as difference ($\bdiff$), symmetric difference ($\bsdiff$), and double implication ($\bdimp$) can be introduced for convenience in the expected way.

We also introduce the infinitary lattice operations $\binf$ and $\bsup$ which operate on sets of sets (of points) and return a set. Thus, they have as type: $(\wsig)\sigma \ar \wsig$, explicitly written as: $((\ww \ar \bool) \ar \bool) \ar (\ww \ar \bool)$.
\begin{align*}
\binf S \equdef\; \lambda w .~\forall X .~(S~X) \imp (X~w) \;\;\;\;\;\;\;\;\;\;\;
\bsup S \equdef\; \lambda w .~\exists X .~(S~X) \land (X~w)
\end{align*}

Observe that in the spirit of the shallow semantical embedding approach \cite{J41,J23}, $\binf$ and $\bsup$ reuse the meta-logical quantifiers $\forall$ and $\exists$ (cf.~\S\ref{subsec:SSE}). It is easy to see (and it has been automatically checked) that they correspond indeed to the lattice-theoretical infimum and supremum operations respectively; i.e.~our encoded Boolean algebras are \textit{complete} as lattices.

Other convenient notions can become easily encoded. For instance, sets of sets (of points) can be said to be closed under (infinitary) meets or joins.
\begin{align*}
\meetcl~S \equdef&~~ \forall X.\,\forall Y.~ (S~X) \land (S~Y) \imp S~(X \band Y) \\
\joincl~S \equdef&~~ \forall X.\,\forall Y.~ (S~X) \land (S~Y) \imp S~(X \bor Y) \\
\infmcl~S \equdef&~~ \forall D.~ D \bprec S \imp (S\;\,{\binf}D) \\
\supmcl~S \equdef&~~ \forall D.~ D \bprec S \imp (S\;\,{\bsup}D) 
\end{align*}
Note that in the latter two definitions the binary relation $\bprec$ is now relating terms of type $(\wsig)\sigma$.\footnote{Recall that our object-logical terms are type-parameterized, e.g., the relation $\bprec$ can be employed for ordering any pair of sets (of sets (of\dots sets (of points))), i.e., terms of type $((\wsig \dots )\sigma)\sigma$.}
We have also encoded variants in which the subset $D$ is assumed non-empty (which behave slightly differently \cite{AFP}).

As our goal is to use HOL-reasoning tools for performing object-logical derivations (after unfolding connectives' definitions), some remarks on the `faithfulness' of our encoding are in order. The first part of the faithfulness question (`soundness') is generally straightforward to resolve by checking that the HOL-encoded axioms and rules of the object logic are (meta-logically) valid. In fact, automated tools can easily be employed to verify this at any time. In the case of the (logic of) Boolean algebras just introduced, this is intuitively evident, since the definitions of the algebraic operations directly reuse HOL's Boolean connectives. The second part of the question (`completeness') involves extra-logical theoretical analysis in order to assess the extent to which the idiosyncratic features of the meta-logic (HOL) have a bearing on the features of the encoded (i.e.~modeled) object-logical semantic structures (e.g., equational classes of algebras).\footnote{Of course, we can simply assume that our object logics are defined semantically via their shallow embeddings (wrt.~HOL as a meta-language). This assumption, by default, resolves any questions regarding `faithfulness' of their encoding.}

In the present case, we can in fact show that the combination of HOL's $\lambda$-abstraction and primitive equality (see \S\ref{subsec:SSE}) allows us to synthesize terms that can play the role of Boolean algebra \textit{atoms}. Observe that our encoding validates the following HOL formula:
$$\forall P .~P \boldsymbol{\not =} \bbot \imp \exists Q .~\texttt{atom}~Q \land Q \bprec P$$
where~ $\texttt{atom}~A \equdef A \boldsymbol{\not =} \bbot~\land~\forall P .~A \bprec P \,\lor\, A \bprec {\bcmpl} P$.

The proof argument is straightforward: just instantiate $Q$ with the term: $\{y\} \equdef (\lambda x.~x = y)$ for some $y$ such that $(P~y)$ holds, i.e.~$Q$ is the `singleton' (aka.~`unit-set') generated by $y$.
Hence Boolean algebras, as encoded in our approach, are by default \textit{atomic} (and \textit{complete} as mentioned previously). In fact, under the `standard' HOL semantics they are just powerset lattices.

In any case, this atomicity restriction does not cause any trouble when it comes to providing semantics for propositional logics based upon (extensions of) Boolean algebras, since, from Stone's representation theorem, we know that every Boolean algebra is a subalgebra of an (atomic) powerset lattice; this entails that if a propositional formula is (counter)satisfiable in a Boolean algebra then it is also (counter)satisfiable in one that is atomic. This way, faithfulness for our HOL-embedded non-classical logics becomes guaranteed at the propositional level in spite of atomicity.

The question of atomicity becomes actually relevant in the case of propositional and higher-order quantification. In this regard, we note that more general classes of Boolean algebras could, in principle, be obtained in HOL-like systems that do not incorporate extensionality or primitive equality.\footnote{The presence of primitive equality in HOL and related systems has become commonplace in the wake of the work of Andrews \cite{andrews1972general}. Automated theorem provers for non-extensional systems are somewhat uncommon, perhaps due to significant limitations (see \cite{andrews1972general,J6} and \cite{B5} for a discussion).} Another approach could be to introduce specialized quantifiers over sets with appropriate restrictions (excluding singletons, and so forth). We will discuss these types of restrictions in \S\ref{subsec:quantifiers}. In any case, since no clear consensus exists on how quantification should behave in non-classical contexts, we restrict ourselves to providing semantically well-motivated definitions for quantifiers in \S\ref{subsec:quantifiers}. We will not touch upon the issue of faithfulness again in this paper.

\subsection{Set-valued Functions and Operators}\label{subsec:operators}

We will refer to those functions whose codomain consists in sets (of points) as \textit{set-valued functions}. They will be encoded as terms having a (both $\itype$- and $\ww$-parametric) type: $\itype \ar \wsig$ (and we will often employ the Greek letters $\vfi$, $\psi$, $\eta$ to denote them).

In the spirit of the shallow semantical embedding approach, we can conveniently define the following ($\itype$-type-lifted) Boolean operations on set-valued functions.
\begin{align*}
\vfi ~\bsqand~ \psi \equdef&~ \lambda x .~(\vfi~x) \band (\psi~x) \;\;\; &
\psi ~\bsqimp~ \vfi \equdef&~ \lambda x .~(\psi~x) \bimp (\vfi~x) \\
\vfi ~\bsqor~ \psi \equdef&~ \lambda x .~(\vfi~x) \bor (\psi~x) \;\;\; &
\supc{\vfi} \equdef&~ \lambda x .~\bcmpl(\vfi~x)
\end{align*}
Hence we obtain a \textit{Boolean algebra of set-valued functions}\footnote{Note that set-valued functions are encoded as curried (cf.~\S\ref{subsec:SSE}) binary relations. Thus, we can easily define (relation) composition and transpose on top of our algebra of set-valued functions to obtain a relation algebra.} featuring meet ($\bsqand$), join ($\bsqor$) and implication ($\bsqimp$) as binary connectives of type $(\itype \ar \sigma)\ar(\itype \ar \sigma)\ar(\itype \ar \sigma)$ and, in particular, \textit{complement} $\supc{(\cdot)}$ as a unary connective of type $(\itype \ar \sigma)\ar(\itype \ar \sigma)$. 

This algebra also features two natural 0-ary connectives (or `constants') of type $\itype \ar \sigma$
$$\bsqtop \equdef~ \lambda x .~\btop \text{\;\;\;\;and\;\;\;\;} \bsqbot \equdef~ \lambda x .~\bbot \,.$$
Clearly, the usual interrelations between the Boolean algebra connectives obtain, so that this algebra could have also been presented by employing, e.g., the minimal signature $\langle \bsqand, \supc{(\cdot)}\rangle$. However, we prefer to give a presentation that explicitly reuses the corresponding connectives at the previous level as we did before. We think that this has a pedagogic merit, and, moreover, we have observed that this practice seems to improve the performance of automated reasoning tools.

\begin{figure}
	\centering
	\includegraphics[width=.75\textwidth]{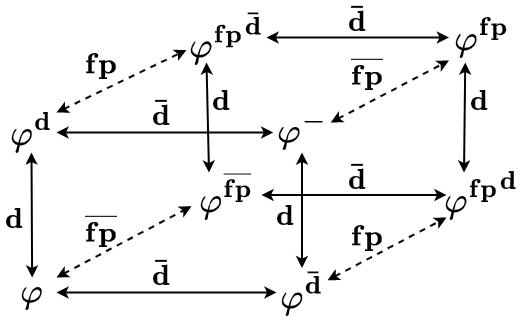}
	\caption{Topological cube of opposition \textit{in abstracto}. Edges denote (involutive) \textit{transformations} on operators, so that the whole diagram commutes. Note that the diagonals, which correspond to complement $\supc{(\cdot)}$, are not shown to avoid visual clutter.}
	\label{fig:cube-abstract}
\end{figure}

Analogous to set-valued functions we can also have \textit{set-domain functions}, i.e., terms of type $\sigma \ar \itype$. For them we can define a convenient unary operation which we call \textit{dual-complement} having type $(\sigma \ar \itype)\ar(\sigma \ar \itype)$, as follows:
$$\supdc{\vfi} \equdef~ \lambda X .~\vfi~({\bcmpl}X) \,.$$
We extend the encoding of Boolean algebras by adding further unary operations (with type $\sigma\ar\sigma$), these we call: \textit{operators}. Moreover, we will often employ an extra (often redundant) pair of parentheses as a suggestive visual aid when applying operators to arguments, so that we will often write $\vfi(A)$ instead of ($\vfi\,A)$.
In fact, operators are both set-valued and set-domain functions at the same time.
Thus, an \textit{algebra of operators} inherits the connectives defined above plus the following one, which we aptly call \textit{dual}:
$$\supd{\vfi} \equdef~ \lambda X .~\bcmpl\vfi(\bcmpl X) \,.$$
Moreover, given an operator $\vfi$, we define the set of its fixed points $(\fp{\vfi})$ as:
$$\fp{\vfi} \equdef \lambda X.~\vfi(X) \bapprox X \,.$$
On top of this we can introduce a further operation $\supfp{(\cdot)}$, dubbed \textit{fixed point}:
$$\supfp{\vfi} \equdef \lambda X.~\vfi(X) \bdimp X \,,$$
noting that the following fundamental interrelation  obtains:
$$(\fp{\vfi})~X \longleftrightarrow \supfp{\vfi}(X) \bapprox \btop \,.$$
We also introduce convenient notation for the complement of the operation $\supfp{(\cdot)}$, which we call \textit{fixed-point-complement}: $\supcfp{\vfi} \equdef \supc{(\supfp{\vfi})} =\supfp{(\supc{\vfi})}$. In fact we have that
$$\supcfp{\vfi} \equ \lambda X.~\vfi(X) \,\bsdiff\, X  \,.$$
We will often refer to the unary operations (on operators) introduced above, namely $\supc{(\cdot)}$, $\supd{(\cdot)}$, $\supdc{(\cdot)}$, $\supfp{(\cdot)}$ and $\supcfp{(\cdot)}$, as \textit{transformations}. We note that they are all involutions, i.e., $(\vfi^f)^f = \vfi$ for $f$ in $\{\texttt{d}, \cmplbar, \dbar, \texttt{fp}, \fpbar \}$. We verified several useful interrelations among them, which are best illustrated by means of a commutative diagram, as shown in Fig.~\ref{fig:cube-abstract}, which will be instantiated later more concretely as a \textit{topological cube of opposition}.

The \textit{algebra of operators} just introduced can be given in fact a minimal signature: $\langle \bn, \circ, \bsqand \rangle$, where $\bn$ is a 0-ary connective, defined in the present work as the Boolean complement, i.e., $\bn \equdef \bcmpl$,\footnote{Future work shall explore the consequences of weakening this definition, so that $\bn$ can stand for an arbitrary involution.} and $\circ$ is an associative binary connective denoting function (operator) composition. We begin by introducing the 0-ary connective $\be \equdef \bn \,\circ\, \bn$, which corresponds in fact to the identity function ($\lambda X.~X$). Thus $\langle\be,\circ\rangle$ is a monoid. We show below how to define the other connectives in terms of $\bn$, $\be$, $\circ$, and $\bsqand$:
\begin{align*}
	\supc{\vfi} &\equdef \bn \,\circ\, \vfi \\
	\supdc{\vfi} &\equdef \vfi \,\circ\, \bn \\
	\supd{\vfi} &\equdef \bn \,\circ\, \vfi \,\circ\, \bn \\
	\vfi ~\bsqor~ \psi &\equdef \supc{(\supc{\vfi} ~\bsqand~ \supc{\psi})} \\
	\bsqtop &\equdef \bn ~\bsqor~ \be \\
	\bsqbot &\equdef \bn ~\bsqand~ \be \\
	\supfp{\vfi} &\equdef (\vfi ~\bsqand~ \be) ~\bsqor~ (\supc{\vfi} ~\bsqand~ \bn) \\
	\supcfp{\vfi} &\equdef (\vfi ~\bsqor~ \be) ~\bsqand~ (\supc{\vfi} ~\bsqor~ \bn) \,.
\end{align*}

\subsection{Topological Conditions on Operators}\label{subsec:conditions}

Recalling from \S\ref{sec:preliminaries} the axiomatic conditions stated in Def.~\ref{def:closure} and Def.~\ref{def:interior}, corresponding to the Kuratowski closure conditions and their duals, we shall now treat those as properties of unary operations (operators), which we may thus call (for lack of better names): \textit{additive} (\ADDI),  \textit{multiplicative} (\MULT), \textit{expansive} (\EXPN), \textit{contractive} (\CNTR), \textit{normal} (\NORM), \textit{dual-normal} (\DNRM), or \textit{idempotent} (\IDEM). We shall also suggestively refer to them as a whole as \textit{positive} properties or conditions. We present in Table~\ref{table:conditions-duals} the corresponding definitions as encoded in HOL. For the sake of finer granularity, definitions involving equalities have been stated separately for the $\bprec$ direction and for the $\bsucc$ direction, conveniently using the suffixes `$a$' and `$b$' to facilitate reference.\footnote{Thus $\ADDI = \ADDI^a \,\land\, \ADDI^b$, $\MULT = \MULT^a \,\land\, \MULT^b$, and $\IDEM = \IDEM^a \land \IDEM^b$. Also observe that $\IDEM^b$ and $\IDEM^a$ are instances of \EXPN\ and \CNTR\ respectively (therefore we omit them sometimes). Also note that to avoid visual clutter we don't display universal quantifiers in definitions, and thus free variables ($A$, $B$, etc.) shall always be read as being universally quantified.}

\begin{table}\centering
	\begin{tabular}{ |c|c| } 
		\hline			
		\textbf{condition (for any $\vfi$)} & \textbf{dual condition (for $\psi = \supd{\vfi}$)}  \\ 
		\hline
		$\texttt{ADDI}^a~\vfi \equdef \vfi(A\bor B) \bprec \vfi(A) \bor \vfi(B)$ & $\texttt{MULT}^b~\psi \equdef \psi(A\band B) \bsucc \psi(A) \band \psi(B)$ \\
		$\texttt{ADDI}^b~\vfi \equdef \vfi(A\bor B) \bsucc \vfi(A) \bor \vfi(B)$ & $\texttt{MULT}^a~\psi \equdef \psi(A\band B) \bprec \psi(A) \band \psi(B)$ \\ 
		$\EXPN~\vfi \equdef A \bprec \vfi(A)$ & $\CNTR~\psi \equdef \psi(A) \bprec A$ \\ 
		$\NORM~\vfi \equdef \vfi(\bbot) \bapprox \bbot$ & $\DNRM~\psi \equdef \psi(\btop) \bapprox \btop$ \\ 
		$\texttt{IDEM}^a~\vfi \equdef \vfi(\vfi(A)) \bprec \vfi(A)$ & $\texttt{IDEM}^b~\psi \equdef \psi(A) \bprec \psi(\psi(A))$ \\ 
		\hline
	\end{tabular}
	\caption{Some \textit{positive} properties of operators (Kuratowski conditions and their duals).}
	\label{table:conditions-duals}
\end{table}

The conditions presented above are not fully independent. We have in fact that the conditions $\ADDI^b$ and $\MULT^a$ are equivalent. They are actually equivalent to monotonicity (\MONO), a fundamental property of operators that deserves special mention:
$$\MONO~\vfi \equdef A \bprec B \imp \vfi(A) \bprec \vfi(B) \,.$$
Thus, we have that $\MONO~\vfi \text{\;iff\;} \ADDI^b~\vfi \text{\;iff\;} \MULT^a~\vfi$. In fact, monotonicity is self-dual: $\MONO~\vfi \text{\;iff\;} \MONO~\supd{\vfi}$.
Clearly, $\EXPN\,\vfi$ implies both $\DNRM\,\vfi$ and $\IDEM^b\,\vfi$, and $\CNTR\,\vfi$ implies both $\NORM\,\vfi$ and $\IDEM^a\,\vfi$.

The infinitary generalizations for \ADDI\ resp.~\MULT, referred to as \iADDI\ resp.~\iMULT\ (cf.~Table~\ref{table:conditions-duals-infinitary}) play an important role in connecting our results to the semantics of normal modal logics, i.e., modal logics characterizable by means of relational (Kripke) frames. We start by noting that, besides entailing \ADDI\ resp.~\MULT, they also entail \NORM\ resp.~\DNRM, since the latter can be seen as corresponding to distribution over empty suprema resp.~infima.

\begin{table}\centering
	\begin{tabular}{ |c|c| } 
		\hline			
		\textbf{condition (for any $\vfi$)} & \textbf{dual condition (for $\psi = \supd{\vfi}$)}  \\ 
		\hline
		$\texttt{iADDI}^a~\vfi \equdef \vfi(\bsup S) \bprec \bsup \llbracket \vfi~S \rrbracket$ & $\texttt{iMULT}^b~\psi \equdef \psi(\binf S) \bsucc \binf \llbracket \psi~S \rrbracket$ \\
		$\texttt{iADDI}^b~\vfi \equdef \vfi(\bsup S) \bsucc \bsup \llbracket \vfi~S \rrbracket$ & $\texttt{iMULT}^a~\psi \equdef \psi(\binf S) \bprec \binf \llbracket \psi~S \rrbracket$ \\ 
		\hline
	\end{tabular}
	\caption{Infinitary \textit{positive} conditions for \ADDI\ (left) and their dual counterparts (right). 
	We employ the notation $\llbracket\,\cdot~\,\cdot\,\rrbracket$ to denote the image of a set under a function, i.e., $\llbracket f~S \rrbracket \equdef \lambda y.~\exists x.~(S~x)\land~(f~x) = y$.}
	\label{table:conditions-duals-infinitary}
\end{table}

In topological terms, closure operators satisfying \iADDI\ (together with the other Kuratowski conditions \EXPN\ and \IDEM) are precisely those giving rise to so-called Alexandrov (or `finitely generated') topologies.\footnote{Here we shall remark that \iADDI\ resp.~\iMULT\ are strictly stronger conditions than \ADDI\ resp.~\MULT\ (e.g., they entail \NORM\ resp.~\DNRM); they are directly satisfied by closure resp.~interior operators defined on relational frames in the usual way (i.e.~as the modal operators $\Diamond$ and $\Box$ respectively). Alexandrov spaces can be alternatively characterized as those in which the intersection (union) of an arbitrary family of open (closed) sets is open (closed). In algebraic terms, this means that the set of fixed points of the interior (closure) operator is closed under infinite meets (joins). Of course, both characterizations are equivalent when assuming all Kuratowski conditions. This is, however, not the case for `weaker' axiomatizations. The interested reader is referred to the formalization sources \cite{AFP}, where we investigate under which \textit{minimal} conditions these results (among others) hold.} It is well known that every Alexandrov topology corresponds, in a sense, to a reflexive and transitive relation (its so-called \textit{specialization preorder}). This correspondence can actually be generalized by factoring out Kuratowski conditions other than \iADDI; in fact, the corresponding relation does not need to be a preorder (we may thus refer to it as a \textit{reachability relation}).\footnote{It is worth mentioning that in Alexandrov topologies every point has a minimal/smallest neighborhood, namely the set of points related to it by the specialization preorder (or, more generally, the reachability relation).}
More specifically, the following HOL formula is valid:
$$ \iADDI~\vfi \dimp \vfi = \C[\R[\vfi]] \,,$$
where $\C[R]$ and $\R[\vfi]$ transform, respectively, a relation into a closure(-like) operator and an operator $\vfi$ (intended as closure) into a (reachability) relation, i.e.,
$$\C[R] \equdef \lambda A.~\lambda w.~\exists v.~R~w~v \land A~v  \text{\;\;\;conversely\;\;\;} \R[\vfi] \equdef \lambda w.~\lambda v.~\vfi~\{v\}~w \,.$$

Let $\psi = \C[R]$ for an arbitrary relation $R$. We have thus that $\iADDI~\psi$ and $\NORM~\psi$ always hold. Moreover, we shall observe that the usual correspondences obtain, as known from modal logic: $\EXPN~\psi$ iff $R$ is reflexive, $\IDEM^a~\psi$ iff $R$ is transitive, and so on.  Furthermore, the relation $\R[\vfi]$ will be reflexive if $\EXPN~\vfi$ holds, and transitive if both $\MONO~\vfi$ and $\IDEM^a~\vfi$ hold (see \cite{AFP}).

In fact we have that the conditions $\iADDI^b$ and $\iMULT^a$ are equivalent to their finitary versions, and thus $\MONO$, $\ADDI^b$, $\MULT^a$, $\iADDI^b$ and $\iMULT^a$ are all equivalent.  
Interesting interrelations with fixed points also obtain:
 $\MULT\,\vfi$ implies~ $\meetcl\,(\fp{\vfi})$ and $\iMULT\,\vfi$ implies $\infmcl\,(\fp{\vfi})$, while the converse holds under the additional assumptions $\MONO\,\vfi$, $\CNTR\,\vfi$ and $\IDEM^b\,\vfi$;
$\ADDI\,\vfi$ implies~ $\joincl\,(\fp{\vfi})$ and $\iADDI\,\vfi$ implies $\supmcl\,(\fp{\vfi})$, while the converse holds under the additional assumptions $\MONO\,\vfi$, $\EXPN\,\vfi$ and $\IDEM^a\,\vfi$.
Moreover, assuming $\MONO\,\vfi$, we have that $\EXPN\,\vfi$ implies $\infmcl\,(\fp{\vfi})$ and that $\CNTR\,\vfi$ implies $\supmcl\,(\fp{\vfi})$.
We refer the interested reader to the Isabelle/HOL sources \cite{AFP} for more results.

We have just seen how the conditions above can be presented in `dual pairs', meaning that when an operator $\vfi$ satisfies one condition, its dual $\supd{\vfi}$ satisfies its dual condition. Another interesting set of `dual pairs' of conditions is listed in Table~\ref{table:conditions-compl}. These are the conditions satisfied by the \textit{complement} and \textit{dual-complement} of closure operators, to which we suggestively refer as \textit{negative} conditions (whereby the previous ones are their \textit{positive} counterparts).

\begin{table}\centering
\begin{tabular}{ |c|c|c| } 
\hline			
	\textbf{ref.~(any $\chi$)} & \textbf{complement cond.~(for $\vfi = \supc{\chi}$)} & \textbf{dual-complement cond.~(for $\psi = \supd{\vfi} = \supdc{\chi}$)}  \\ 
	\hline
	$\texttt{ADDI}^a~\chi$ & $\texttt{nADDI}^a~\vfi \equdef \vfi(A\bor B) \bsucc \vfi(A) \band \vfi(B)$ & $\texttt{nMULT}^b~\psi \equdef \psi(A\band B) \bprec \psi(A) \bor \vfi(B)$\\
	$\texttt{ADDI}^b~\chi$ & $\texttt{nADDI}^b~\vfi \equdef \vfi(A\bor B) \bprec \vfi(A) \band \vfi(B)$ & $\texttt{nMULT}^a~\psi \equdef \psi(A\band B) \bsucc \psi(A) \bor \psi(B)$\\
	$\texttt{iADDI}^a~\chi$ & $\texttt{inADDI}^a~\vfi \equdef \vfi(\bsup S) \bsucc \binf \llbracket \vfi~S \rrbracket$ & $\texttt{inMULT}^b~\psi \equdef \psi(\binf S) \bprec \bsup \llbracket \psi~S \rrbracket$ \\
	$\texttt{iADDI}^b~\chi$ & $\texttt{inADDI}^b~\vfi \equdef \vfi(\bsup S) \bprec \binf \llbracket \vfi~S \rrbracket$ & $\texttt{inMULT}^a~\psi \equdef \psi(\binf S) \bsucc \bsup \llbracket \psi~S \rrbracket$ \\ 
	$\EXPN~\chi$ & $\texttt{nEXPN}~\vfi \equdef \vfi(A) \bprec \bcmpl A$ & $\texttt{nCNTR}~\psi \equdef \bcmpl A \bprec \psi(A)$ \\ 
	$\NORM~\chi$ & $\texttt{nNORM}~\vfi \equdef \vfi(\bbot) \bapprox \btop$ & $\texttt{nDNRM}~\psi \equdef \psi(\btop) \bapprox \bbot$\\ 
	$\texttt{IDEM}^a~\chi$ & $\texttt{nIDEM}^a~\vfi \equdef \vfi(A) \bprec \vfi(\bcmpl\vfi(A))$ & $\texttt{nIDEM}^b~\psi \equdef \psi(\bcmpl\psi(A)) \bprec \psi(A)$\\ 
\hline
\end{tabular}
\caption{Some \textit{negative} conditions (taking closure conditions as reference).}
\label{table:conditions-compl}
\end{table}

The prefix `\texttt{n}' in the names of the conditions in Table \ref{table:conditions-compl} can suggestively be read as `\textit{anti}', and thus, e.g., \texttt{nADDI}, \texttt{nEXPN} and \texttt{nIDEM} could be read as \textit{anti-additivity}, \textit{anti-expansiveness} and \textit{anti-idempotence}, respectively. We can also define the negative counterpart to the monotonicity condition, dubbed \textit{antitonicity} or \textit{anti-monotonicity} (\texttt{ANTI}):
$$\texttt{ANTI}~\vfi \equdef A \bprec B \imp \vfi(B) \bprec \vfi(A) \,.$$

Unsurprisingly, \texttt{ANTI} is self-dual. In fact, analogously as before, $\texttt{ANTI}$, $\texttt{nADDI}^b$, $\texttt{nMULT}^a$, $\texttt{inADDI}^b$, and $\texttt{inMULT}^a$ end up being all equivalent. Analogous interrelations obtain here as in the previous case, replacing the conditions by their complement (negative) counterparts (see sources \cite{AFP}).

\begin{figure}
	\centering
	\includegraphics[width=.55\textwidth]{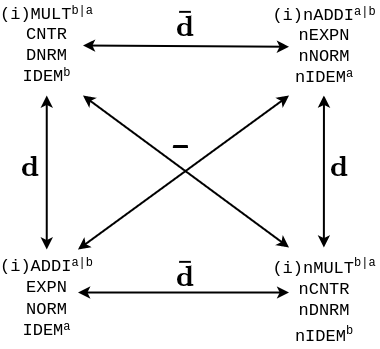}
	\caption{Interrelations between some positive and negative conditions for operators.}
	\label{fig:square-conditions1}
\end{figure}

In fact, the interrelations between the axiomatic conditions satisfied by an operator $\vfi$, its dual $\supd{\vfi}$, its complement $\supc{\vfi}$ and its dual-complement $\supdc{\vfi}$ (Tables \ref{table:conditions-duals}, \ref{table:conditions-duals-infinitary} and \ref{table:conditions-compl}), can also be conveniently represented in a kind of \textit{square of opposition}, as illustrated in Fig.~\ref{fig:square-conditions1}.

In the same spirit, we shall now introduce pairs of related conditions corresponding to the fixed-point transformations $\supfp{(\cdot)}$ and $\supcfp{(\cdot)}$. However, our computer-supported experiments \cite{AFP} quickly led us to realize that the axiomatic conditions discussed so far are too coarse-grained to allow for establishing one-to-one correspondences (as in Tables \ref{table:conditions-duals}-\ref{table:conditions-compl}). In fact, the axiomatic conditions need to be considered in what we call a `relativized' form.

Broadly speaking, relativizing an axiomatic condition, in our sense, consists in restricting the domain of points (in equalities or operators) wrt.~a `dynamically' constructed set. This `relativizing' set is thus not fixed but defined in terms of the (universally quantified) variables featured in the definition. As an illustration, the condition
$$\texttt{nMULT}~\psi \equdef \psi (A\band B) ~\bapprox \psi(A) \bor \psi(B) \,,$$
which is equivalently stated (with explicit quantification over points) as
$$\texttt{nMULT}~\psi \equdef \forall w.~{\psi (A\band B)~w \dimp (\psi(A) \bor \psi(B))~w} \,,$$
can be relativized as
$$\texttt{nMULTr}~\psi \equdef \forall w.~{(A \band B)~w} \imp (\psi (A\band B)~w \dimp (\psi(A) \bor \psi(B))~w) \,,$$
which is, in fact, equivalently formulated as 
$$\texttt{nMULTr}~\psi \equdef {(A \band B)~\band~} \psi (A\band B) ~\bapprox {(A \band B)~\band~} (\psi(A) \bor \psi(B)) \,.$$

We take the opportunity to observe that \texttt{nMULTr} above corresponds to the axiomatic condition \texttt{B1} for topological \textit{border} operators, as presented in Def.~\ref{def:border} in \S\ref{sec:preliminaries}. This is of course not a coincidence, as will be discussed in \S\ref{subsec:cube-concrete}.

We shall now introduce a convenient notation for (in)equalities relativized  wrt.~the points inside resp.~outside of a given set $U$
\begin{align*}
	A\;{\bapprox_U}\,B &\equdef\;\forall x.~~\,U\,x \imp (A\,x \dimp B\,x)~~~~~~~{A\,{\bprec_U}\,B} &\equdef\;\; &\forall x.~~\,U\,x \imp (A\,x \imp B\,x)\\
	A\;{\bapprox^U}\,B &\equdef\;\forall x.\,\neg U\,x \imp (A\,x \dimp B\,x)~~~~~~~A\,{\bprec^U}\,B &\equdef\;\; &\forall x.\,\neg U\,x \imp (A\,x \imp B\,x)
\end{align*}
which we shall call the \textit{lower} resp. \textit{upper} relativization (of a given (in)equality) wrt.~ an element $U$ (in our case a set).
Observe that the definitions above have alternative equivalent formulations as
\begin{align*}
	A \bapprox_U B &\equ (U \band A \bapprox U \band B)~~~~~~~~~~~~& A \bprec_U B &\equ (U \band A \bprec U \band B) \\
	A \bapprox^U B &\equ (U \bor A \bapprox U \bor B)~~~~~~~~~~~~& A \bprec^U B &\equ (U \bor A \bprec U \bor B) \,.
\end{align*}
These definitions above can thus be interpreted as the relativization of (in)equalities wrt.~the elements below resp. above $U$; or, algebraically speaking, wrt.\ the principal ideal resp.\ filter generated by $U$.

We now list the corresponding relativized variants for axiomatic conditions. We begin by noting that, in our case, the relativizing terms can at most feature only \textit{bound} variables (see e.g.~$(A \band B)$ in \texttt{nMULTr} above). As a consequence, relativization cannot be meaningfully applied to conditions \texttt{(n)NORM} or \texttt{(n)DNRM}. Similarly, relativizing the conditions \texttt{(n)EXPN} or \texttt{(n)CNTR} either leaves them unchanged or trivializes them.
Thus, relativization meaningfully applies only to additivity and idempotence and derived conditions. We show in Tables~\ref{table:conditions-duals-rel} and \ref{table:conditions-compl-rel} the \textit{appropriate} (for our purposes) relativized variants of the positive and negative axiomatic conditions previously discussed.\footnote{Observe that when iterating operations (as in \texttt{(n)IDEM}) the relativization of their domain becomes evident. These relativized variants have been elicited with the crucial help of automated tools, in particular Isabelle's \textit{Sledgehammer} \cite{blanchette2016hammering} and \textit{Nitpick} \cite{Nitpick}.}

\begin{table}\centering
	\begin{tabular}{ |c|c| } 
		\hline			
		\textbf{relativized condition (for any $\vfi$)} & \textbf{relativized dual condition (for $\psi = \supd{\vfi}$)}  \\ 
		\hline
		$\texttt{ADDIr}^a~\vfi \equdef \vfi(A\bor B) \bprec^{A\bor B} \vfi(A) \bor \vfi(B)$ & $\texttt{MULTr}^b~\psi \equdef \psi(A\band B) \bsucc_{A\band B} \psi(A) \band \psi(B)$ \\
		$\texttt{ADDIr}^b~\vfi \equdef \vfi(A\bor B) \bsucc^{A\bor B} \vfi(A) \bor \vfi(B)$ & $\texttt{MULTr}^a~\psi \equdef \psi(A\band B) \bprec_{A\band B} \psi(A) \band \psi(B)$ \\
		$\texttt{iADDIr}^a~\vfi \equdef \vfi(\bsup S) \bprec^{\bsup S} \bsup \llbracket \vfi~S \rrbracket$ & $\texttt{iMULTr}^b~\psi \equdef \psi(\binf S) \bsucc_{\binf S} \binf \llbracket \psi~S \rrbracket$ \\
		$\texttt{iADDIr}^b~\vfi \equdef \vfi(\bsup S) \bsucc^{\bsup S} \bsup \llbracket \vfi~S \rrbracket$ & $\texttt{iMULTr}^a~\psi \equdef \psi(\binf S) \bprec_{\binf S} \binf \llbracket \psi~S \rrbracket$ \\	
		$\texttt{IDEMr}^a~\vfi \equdef \vfi(A\bor \vfi(A)) \bprec^{A} \vfi(A)$ & $\texttt{IDEMr}^b~\psi \equdef \psi(A) \bprec_{A} \psi(A\band \psi(A))$ \\ 
		\hline
	\end{tabular}
	\caption{Relativized variants for some \textit{positive} properties of operators (cf.~Tables \ref{table:conditions-duals}-\ref{table:conditions-duals-infinitary}).}
	\label{table:conditions-duals-rel}
\end{table}
\begin{table}\centering
	\begin{tabular}{ |c|c| } 
		\hline			
		\textbf{relativized condition (for any $\vfi$)} & \textbf{relativized dual condition (for $\psi = \supd{\vfi}$)}  \\ 
		\hline
		$\texttt{nADDIr}^a~\vfi \equdef \vfi(A\bor B) \bsucc^{A\bor B} \vfi(A) \band \vfi(B)$ & $\texttt{nMULT}^b~\psi \equdef \psi(A\band B) \bprec_{A\band B} \psi(A) \bor \vfi(B)$\\
		$\texttt{nADDIr}^b~\vfi \equdef \vfi(A\bor B) \bprec^{A\bor B} \vfi(A) \band \vfi(B)$ & $\texttt{nMULT}^a~\psi \equdef \psi(A\band B) \bsucc_{A\band B} \psi(A) \bor \psi(B)$\\
		$\texttt{inADDIr}^a~\vfi \equdef \vfi(\bsup S) \bsucc^{\bsup S} \binf \llbracket \vfi~S \rrbracket$ & $\texttt{inMULTr}^b~\psi \equdef \psi(\binf S) \bprec_{\binf S} \bsup \llbracket \psi~S \rrbracket$ \\
		$\texttt{inADDIr}^b~\vfi \equdef \vfi(\bsup S) \bprec^{\bsup S} \binf \llbracket \vfi~S \rrbracket$ & $\texttt{inMULTr}^a~\psi \equdef \psi(\binf S) \bsucc_{\binf S} \bsup \llbracket \psi~S \rrbracket$ \\ 
		$\texttt{nIDEMr}^a~\vfi \equdef \vfi(A) \bprec^{A} \vfi(A\bor \bcmpl\vfi(A))$ & $\texttt{nIDEMr}^b~\psi \equdef \psi(A \band \bcmpl\psi(A)) \bprec_{A} \psi(A)$\\ 
		\hline
	\end{tabular}
	\caption{Relativized variants for some \textit{negative} properties of operators (cf.~Table \ref{table:conditions-compl}).}
	\label{table:conditions-compl-rel}
\end{table}

It is worth to have a look at the two pairs of dual conditions below, corresponding to weak variants of \MONO\ and \texttt{ANTI}.
\begin{align*}
	\texttt{MONOw}^1~\vfi \equdef A \bprec B \imp \vfi(A) \bprec B \bor \vfi(B) \\
	\texttt{MONOw}^2~\vfi \equdef A \bprec B \imp A \band \vfi(A) \bprec \vfi(B) \\
	\texttt{ANTIw}^1~\vfi \equdef A \bprec B \imp \vfi(B) \bprec B \bor \vfi(A)  \\
	\texttt{ANTIw}^2~\vfi \equdef A \bprec B \imp A \band \vfi(B) \bprec \vfi(A)
\end{align*}

Unsurprisingly, $\texttt{MONOw}^1$ and $\texttt{MONOw}^2$ are each other's duals. The same holds for $\texttt{ANTIw}^1$ and $\texttt{ANTIw}^2$. In fact, analogously as before, $\texttt{MONOw}^1$ is equivalent to $\texttt{ADDIr}^b$ and $\texttt{MONOw}^2$ is equivalent to $\texttt{MULTr}^a$. Similarly, $\texttt{ANTIw}^1$ is equivalent to $\texttt{nADDIr}^b$ and $\texttt{ANTIw}^2$ is equivalent to $\texttt{nMULTr}^a$ (see sources \cite{AFP} for more interrelations).

The conditions \texttt{(n)EXPN} and \texttt{(n)CNTR} play a fundamental role in relating axiomatic conditions with their relativized variants. Thus,
\begin{itemize}
\item given any of \EXPN\ or \texttt{nEXPN}, we have that $\texttt{ADDI}^a$ and $\texttt{ADDIr}^a$ are equivalent;
\item given \EXPN, we have that $\texttt{ADDI}^b$ and $\texttt{ADDIr}^b$ (and $\texttt{MONOw}^1$) are equivalent;
\item given \CNTR, we have that $\texttt{MULT}^a$ and $\texttt{MULTr}^a$ (and $\texttt{MONOw}^2$) are equivalent;
\item given any of \CNTR\ or \texttt{nCNTR}, we have that $\texttt{MULT}^b$ and $\texttt{MULTr}^b$ are equivalent;
\item given any of \EXPN\ or \texttt{nEXPN}, we have that $\texttt{nADDI}^a$ and $\texttt{nADDIr}^a$ are equivalent;
\item given \texttt{nEXPN}, we have that $\texttt{nADDI}^b$ and $\texttt{nADDIr}^b$ (and $\texttt{ANTIw}^1$) are equivalent;
\item given \texttt{nCNTR}, we have that $\texttt{nMULT}^a$ and $\texttt{nMULTr}^a$ (and $\texttt{ANTIw}^2$) are equivalent;
\item given any of \CNTR\ or \texttt{nCNTR}, we have that $\texttt{nMULT}^b$ and $\texttt{nMULTr}^b$ are equivalent.
\end{itemize}

Having discussed relativized axiomatic conditions and their interrelations, we can now return to the question of establishing one-to-one correspondences between conditions and their fixed-point transformations. First, observe that
\begin{align*}
\EXPN~\vfi \text{\;\;\;iff\;\;\;\;} \EXPN~\supfp{\vfi} &\text{\;\;\;iff\;\;\;} \texttt{nEXPN}~\supcfp{\vfi} \\ 
\CNTR~\vfi \text{\;\;\;iff\;\;} \texttt{nCNTR}~\supfp{\vfi} &\text{\;\;\;iff\;\;\;} \CNTR~\supcfp{\vfi}
\end{align*}
and also
\begin{align*}
\NORM~\vfi \text{\;\;iff\;\;\;\;} \texttt{nNORM}~\supfp{\vfi} &\text{\;\;\;iff\;\;\;} \NORM~\supcfp{\vfi} \\ 
\DNRM~\vfi \text{\;\;\;iff\;\;\;\;} \DNRM~\supfp{\vfi} &\text{\;\;\;iff\;\;\;} \texttt{nDNRM}~\supcfp{\vfi} \,.
\end{align*}

As for relativized variants of additivity and derived negative conditions we have that
\begin{align*}
	\texttt{(i)ADDIr}^a~\vfi \text{\;\;\;iff\;\;\;\;} \texttt{(i)nADDIr}^a~\supfp{\vfi} &\text{\;\;\;iff\;\;\;} \texttt{(i)ADDIr}^a~\supcfp{\vfi} \\ 
	\texttt{(i)ADDIr}^b~\vfi \text{\;\;\;iff\;\;} \texttt{(i)nADDIr}^b~\supfp{\vfi} &\text{\;\;\;iff\;\;\;} \texttt{(i)ADDIr}^b~\supcfp{\vfi}
\end{align*}
and similarly for multiplicativity and derived negative conditions
\begin{align*}
	\texttt{(i)MULTr}^a~\vfi \text{\;\;\;iff\;\;\;\;} \texttt{(i)MULTr}^a~\supfp{\vfi} &\text{\;\;\;iff\;\;\;} \texttt{(i)nMULTr}^a~\supcfp{\vfi} \\ 
	\texttt{(i)MULTr}^b~\vfi \text{\;\;\;iff\;\;} \texttt{(i)MULTr}^b~\supfp{\vfi} &\text{\;\;\;iff\;\;\;} \texttt{(i)nMULTr}^b~\supcfp{\vfi} \,.
\end{align*}
Finally, in the case of \IDEM\ and derived negative conditions we have that
\begin{align*}
	\texttt{IDEMr}^a~\vfi \text{\;\;\;iff\;\;\;\;} \texttt{nIDEMr}^a~\supfp{\vfi} &\text{\;\;\;iff\;\;\;} \texttt{IDEMr}^a~\supcfp{\vfi} \\ 
	\texttt{IDEMr}^b~\vfi \text{\;\;\;iff\;\;} \texttt{IDEMr}^b~\supfp{\vfi} &\text{\;\;\;iff\;\;\;} \texttt{nIDEMr}^b~\supcfp{\vfi} \,.
\end{align*}

\begin{figure}
	\centering
			\includegraphics[width=.99\textwidth]{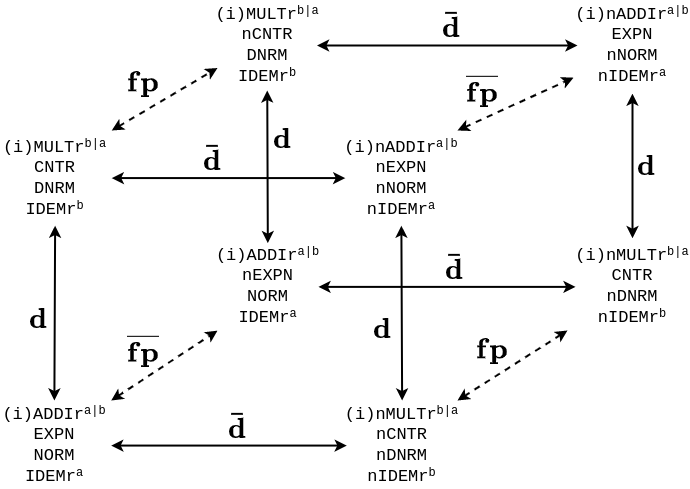}
		\caption{Interrelations between some axiomatic conditions for operators. The lower left corner corresponds to the Kuratowski closure conditions.}
		\label{fig:cube-conditions}
	\end{figure}

In Fig.~\ref{fig:cube-conditions} we show a suggestive representation of the interrelations between the axiomatic conditions discussed so far.

\subsection{A Topological Cube of Opposition}\label{subsec:cube-concrete}

\begin{figure}
	\centering
	\includegraphics[width=.50\textwidth]{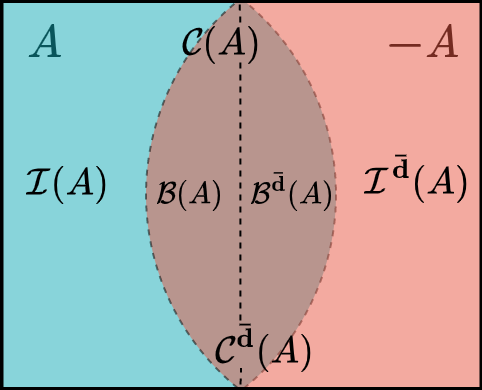}		
	\caption{Suggestive pictorial representation of topological operators and their inter-definitions. Note that the illustration implicitly assumes the second Kuratowski condition (i.e., \EXPN\ for $\C$ and its counterpart for the other operators).}
	\label{fig:topo-operators2}
\end{figure}

We have previously remarked that the four conditions: \ADDI, \EXPN, \NORM\ and \IDEM, correspond to the well-known Kuratowski conditions on \textit{closure} operators (recall \S\ref{sec:preliminaries}~Def.~\ref{def:closure}),
and that any operator $\C$ satisfying some `meaningful' subset of those (possibly weakened) conditions shall be called a \textit{closure operator}.\footnote{As noted before, we are abusing terminology, as we do not fix a minimal set of conditions that an operator needs to satisfy to deserve being called a `closure'; e.g., Moore (hull) closures only satisfy \MONO, \EXPN, and \IDEM, while \v{C}ech closure operators may not satisfy \IDEM\ (they are sometimes called `preclosures'). Anyhow, we let the context dictate how operators are called.}
Similarly, any operator $\I$ partially satisfying the dual conditions (\MULT, \CNTR, \DNRM, \IDEM) shall be called an \textit{interior operator} (cf.~\S\ref{sec:preliminaries}~Def.~\ref{def:interior}). Recall, moreover, that we always obtain an interior operator as the dual of a closure: $\supd{\C} = \I$. We recall from \S\ref{sec:preliminaries} that operators in TBAs are inter-definable (see Table~\ref{table:op-interrelations2} and  Fig.~\ref{fig:topo-operators2}).

\begin{table}
	\centering
	\large
	\begin{tabular}{ |c|c|c|c|c|} 
		\hline
		\small{define\textbackslash using} & $\C$ & $\I$ & $\E$ & $\B$ \\ 
		\hline
		$\C(A)$ & -- & $\supd{\I}(A)$ & $\cmpl\E(A)$ & $A \lor \B({\cmpl}A)$ \\
		\hline
		$\I(A)$ & $\supd{\C}(A)$ & -- & $\E(\cmpl A)$ & $A \land {\bcmpl}\B(A)$ \\
		\hline	
		$\E(A)$ & ${\cmpl}\C(A)$ & $\I({\cmpl}A)$ & -- & ${\cmpl}A \land \supd{\B}(A)$ \\
		\hline		
		$\B(A)$ & $A \land \C({\cmpl}A)$ & $A \land {\cmpl}\I(A)$ & $A \land \supd{\E}(A)$ & -- \\
		\hline
	\end{tabular}
	\caption{Inter-definitions between operators (extract from Table~\ref{table:op-interrelations}). Recall that entries involving border ($\B$) require assuming the corresponding second Kuratowski condition (i.e., \EXPN\ for $\C$, \CNTR\ for $\I$, and so on; cf.~\S\ref{subsec:prelim-TBA}).}
	\label{table:op-interrelations2}
\end{table}

The reader may have noticed that the negative conditions for $\supc{(\cdot)}$ presented in Table~\ref{table:conditions-compl} correspond indeed to the conditions axiomatizing an exterior operator (cf.\S\ref{sec:preliminaries}~Def.~\ref{def:exterior}). Of course, this is not surprising if we recall the inter-definitions between topological operators in Table~\ref{table:op-interrelations2}.

\begin{figure}
	\centering
	\includegraphics[width=.40\textwidth]{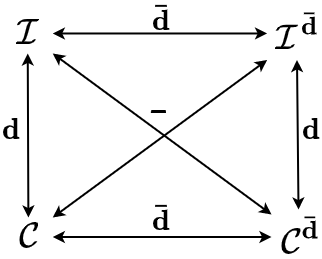}
	\caption{Frontal face of the \textit{topological cube of opposition} (instantiated with $\C$).}
	\label{fig:cube-front}
\end{figure}

\begin{figure}
	\centering
	\includegraphics[width=.67\textwidth]{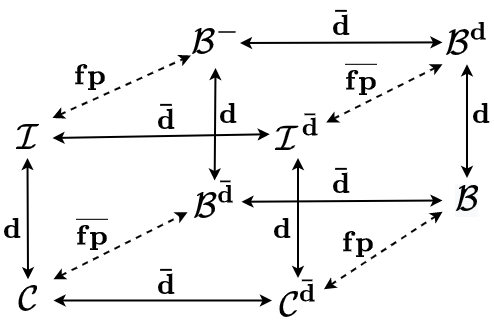}
	\caption{\textit{Topological cube of opposition} instantiated with $\C$ (and $\B = \supd{(\supfp{\C})}$). Note that, for the sake of readability, the diagonals are not shown. They can be easily inferred.}
	\label{fig:cube-concrete}
\end{figure}

Let us now instantiate the commutative diagram shown in Fig.~\ref{fig:cube-abstract} with a closure operator $\C$, and notice that its frontal face corresponds to the \textit{square of opposition} shown in Fig.~\ref{fig:cube-front}, which constitutes itself the frontal face  
of what we suggestively refer to as a \textit{topological cube of opposition} later in Fig.~\ref{fig:cube-concrete}.

\begin{figure}
	\centering
	\includegraphics[width=.40\textwidth]{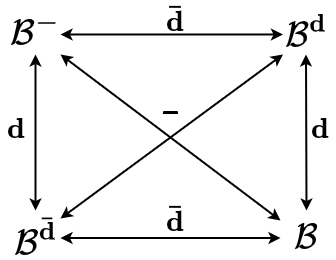}
	\caption{Rear face of the \textit{topological cube of opposition} (instantiated with $\B = \supd{(\supfp{\C})}$).}
	\label{fig:cube-rear}
\end{figure}

Recalling the axiomatic conditions on a \textit{border} operator $\B$ (see \S\ref{sec:preliminaries}~Def.~\ref{def:border}) the reader may also notice, by looking at Fig.~\ref{fig:cube-conditions}, that the conditions satisfied by $\supfp{(\supdc{(\cdot)})}$ (= $\supd{(\supfp{(\cdot)})}$) for a closure operator closely resemble (and are in fact equivalent to\footnote{Observe that the first three conditions coincide. As for the last condition, it can be shown that, in the presence of the other conditions, \texttt{B4} is in fact equivalent to $\texttt{nIDEMr}^b$ (see \cite{AFP}).}) those of a border operator. This is not a coincidence.  In fact a border operator $\B$ can be defined as $\supd{(\supfp{\C})}$ for a given closure operator $\C$. 
Thus, we can notice that the rear face of the commutative diagram in Fig.~\ref{fig:cube-abstract} (instantiated with a closure operator $\C$) gives rise to the \textit{square of opposition} shown in Fig.~\ref{fig:cube-rear}.

Let us now assemble our two squares in Fig.~\ref{fig:cube-front} and Fig.~\ref{fig:cube-rear} into the \textit{topological cube of opposition} in Fig.~\ref{fig:cube-concrete}. 
It is instructive to compare the latter with the corresponding axiomatic conditions in Fig.~\ref{fig:cube-conditions}.

It is also a good exercise to contemplate the cube in Fig.~\ref{fig:cube-concrete} from different perspectives (from above/below, laterally, diagonally, etc.) as it gives rise to quite interesting commutative diagrams. Also interesting is to employ the operations $\bsqand$ and $\bsqor$ to explore symmetries and dualities between pairs of operators, e.g., at opposed corners. As an illustration, we can contemplate the `floor' square and recall that a frontier operator $\F$ is equivalently defined as $\C \,\bsqand\, \supdc{\C}$ and $\supdc{\B} \,\bsqor\, \B$ (cf.~\S\ref{sec:preliminaries}).

\subsection{Non-classical Negations}\label{subsec:negations}

It is well known that closure algebras can serve as semantic structures for modal logic, interpreting the diamond as a closure operator $\C$. By enforcing all the Kuratowski closure conditions on $\C$, this semantics becomes complete for modal logic \textbf{S4}. Naturally, this result can be generalized towards weaker closure(-like) operators. For instance, J\'onsson and Tarski \cite{BAO} demonstrated that by constraining $\C$ with only the conditions \ADDI\ and \NORM\ we obtain a complete semantics for the minimal normal modal logic \textbf{K} (as we call it nowadays). 
Therefore, it is reasonable to think of the (suitably constrained) operator $\C$ and its dual $\supd{\C} = \I$ as \textit{positive modalities}.

On the other hand, we have just seen that other interesting operators are definable from $\C$ by means of the transformations $\supc{(\cdot)}, \supd{(\cdot)}, \supdc{(\cdot)}$, $\supfp{(\cdot)}$ and $\supcfp{(\cdot)}$ as well as the connectives: $\bsqand$, $\bsqor$ and $\bsqimp$. Moreover operators can also be composed as functions. In particular, recalling Fig.~\ref{fig:topo-operators2}, we can see that the operators  $\supdc{\C} = \supc{\I}$ and $\supc{\C} = \supdc{\I} = \E$ make good candidates for paraconsistent and paracomplete negations (\textit{negative modalities}), respectively. Let us define: $\bneg^\C \equdef \supdc{\C}$ and $\bneg^\I \equdef \supdc{\I} = \supd{(\bneg^\C)}$.\footnote{The idea of defining non-classical negations by composing a classical negation with a modality is of course not new and has been proposed in countless occasions by researchers working at the crossroads of modal and non-classical logics (see e.g.~\cite{Marcos2005,Dodo2014} and references therein).}  By employing \textit{degree-preserving consequence} (recall \S\ref{subsec:prelim-logical-consequence}) it is easy to see that `explosion' (ECQ) resp.~`excluded middle' (TND) do not hold for $\bneg^\C$ resp.~$\bneg^\I$, i.e.,
$$ A\,, {\bneg^\C}A \not\vdash \bbot \text{\;\;\;\;and\;\;\;\;} \not\vdash A\,, \bneg^\I A \,.$$

Certainly, for ${\bneg^\C}$ or ${\bneg^\I}$ to behave as a negation worthy of its name, certain conditions must be met by the underlying operator $\C$.\footnote{As one reviewer noted, similar definitions for non-classical negations and related modalities in the literature (e.g., \cite{Marcos2005,Dodo2014,CP17} and references therein) hinge on the validity of \texttt{EXPN} for $\C$ (and its counterparts for other operators: \CNTR\ for $\I$, \texttt{nEXPN} for $\E$, etc.).} The characterization of a `bona fide negation' is a contentious philosophical topic that is orthogonal to the present discussion. We limit ourselves to noting that results such as those presented in \S\ref{subsec:conditions} allow us to `translate' conditions on $\C$ into their counterparts for other operators (in particular $\supc{\C} = {\bneg^\I}$ and $\supdc{\C} = {\bneg^\C}$). This essentially provides a mechanism to control the behavior of potential negations (see, e.g., the properties in Table \ref{table:negation-properties}) by suitably constraining a primitive closure(-like) operator. The interested reader can find in our Isabelle/HOL sources \cite{AFP} a (necessarily partial but steadily growing) list of results in this regard.

\begin{figure}
\centering
\begin{subfigure}{0.49\textwidth}
	\centering
	\includegraphics[width=1\textwidth]{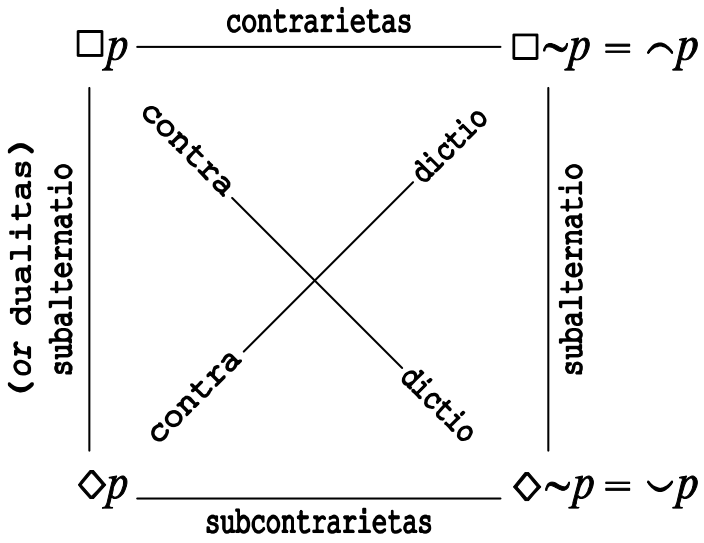}
\end{subfigure}\hfill%
\begin{subfigure}{0.49\textwidth}
\centering
\includegraphics[width=.8\textwidth]{cube-front.png}
\end{subfigure}
\caption{On the left: Marcos' \textit{square of modalities} (excerpt from \cite{Marcos2005}). The connectives $\sim$, $\smallfrown$ and $\smallsmile$ correspond to classical, paracomplete and paraconsistent negation respectively. On the right: front face of the \textit{topological cube of opposition} (as in Fig.~\ref{fig:cube-front}).}
\label{fig:square-modalities}
\end{figure}

It is worth quoting at this point the \textit{square of modalities} as discussed by Marcos \cite{Marcos2005} (and shown in Fig.~\ref{fig:square-modalities}, left), since it will help us set the stage for the discussion on recovery operators in \S\ref{subsec:recovery}. It is worth mentioning that Marcos \cite{Marcos2005} has argued (rightly in our opinion) for a more appropriate interpretation of the vertical edges of the square as ``dualitas'' instead of ``subalternatio''. Observe that this coincides with our previous depiction of the \textit{topological cube of opposition} (see Fig.~\ref{fig:square-modalities}, right).

Finally, we can speculate on the existence of other interesting negation-like connectives obtained by composing operators with each other. Consider, for example, the negation(-like) operators: $\bneg^{\I\C} \equdef \I \,\circ\, \supdc{\C}$ and $\bneg^{\C\I} \equdef \C \,\circ\, \supdc{\I}$ for operators $\C$ and $\I$ (either both being primitive or one defined in terms of the other). Preliminary computer-assisted experiments \cite{AFP} have shown that, by imposing suitable axiomatic (e.g, Kuratowski) conditions on the operators, we can obtain several potentially useful kinds of paradefinite logics in this way. Generally speaking, operations defined according to the schema: $\bneg^{X} \equdef \lambda A. ~g(\C,\bcmpl)(A)$, where $g(\C,\bcmpl)$ is an arbitrary composition of $\C$ and $\bcmpl$, with the latter appearing an odd number of times, are expected to behave as negation(-like) operators. Recalling the limiting result by Kuratowski mentioned in \S\ref{sec:preliminaries}, it follows that we can define at most seven different negations in this manner, restricting ourselves to only one primitive operator (either $\C$ or $\I$). However, keep in mind that Kuratowski's result was obtained for fully constrained (closure) operators. We can therefore anticipate this number to increase as we relax those restrictions. Another viable alternative involves working with several suitably interrelated \textit{primitive} operators (as in the \textit{bi-topological Boolean algebras} studied by Rauszer \cite{rauszer1974semi}). Given the multitude of ways these extensions can be performed, we consider this a potentially interesting topic for future computer-assisted investigations.

\subsection{Recovery Operators}\label{subsec:recovery}

We now shift our focus to the notion of a fixed point as previously discussed. We begin by emphasizing that the fixed-point predicate $(\texttt{fp}~\cdot)$
provides a mechanism for `recovering' many of the classical properties of negation in a sentence-wise fashion. 
 
Regarding the paraconsistent case, we have observed that ${\bneg}^\C$ does not validate ECQ as anticipated.
As it happens, we can `switch on' ECQ for a particular $A$ by simply assuming that $A$ is open, i.e., $\I(A) \bapprox A$ or in other words $(\texttt{fp}~\I)\,A$. Hence, we have
$$(\texttt{fp}~\I)~A \imp A \band {\bneg^\C}A \bapprox \bbot \text{\;\;\;\;i.e.\;if A is open then\;}A\,, {\bneg^\C}A \vdash \bbot \,.$$
 
Dually, when using the paracomplete negation ${\bneg}^\I$, TND can be recovered for a given $A$ by assuming that $A$ is closed. Other `negation-like' semantic conditions (e.g., as in Table \ref{table:negation-properties}) can be similarly recovered by assuming some \textit{minimal} conditions on $\C$. These results have been obtained using Isabelle/HOL; therefore, we direct interested readers to \cite{AFP} for a list of these findings. Furthermore, these results can be easily given an intuitive topological explanation if we recall from \S\ref{sec:preliminaries} that fixed points of operators have interesting alternative characterizations in terms of algebras of sets (e.g., open sets are those with an empty border, and closed sets are those whose complement has an empty border). 

The connection to \textit{object-logical} recovery operators can be established by recalling from \S\ref{subsec:operators} that the fixed-point predicate $(\texttt{fp}~\cdot)$ can be `operationalized' into $\supfp{(\cdot)}$, such that the following holds:
$$(\texttt{fp}~\vfi)~A   ~\dimp~  \supfp{\vfi}(A) \bapprox \btop \,.$$
  
This entails in fact that any `recovery results' obtained by assuming $(\texttt{fp}~\vfi)$, as discussed above, extend directly to $\supfp{\vfi}$. In fact, the following holds in HOL:%
\footnote{Observe that the terms $\Gamma$ and $\Delta$, having type $\sigma \ar \sigma$, can be seen intuitively as corresponding to (arbitrary) formulas with a free variable $A$ (of type $\sigma$). In fact, this HOL-statement can be seamlessly extended to any number of variables.}
$$\supfp{\vfi}(A) \band (\Gamma~A) \bprec (\Delta~A) \imp ((\texttt{fp}~\vfi)\,A \imp (\Gamma~A) \bprec (\Delta~A)) \,.$$
 
As for the paraconsistent case, we note that ECQ (among others) can be recovered by purely object-logical (algebraic) means. Let us introduce the connective ${\bcirc} \equdef \I^\texttt{fp} \;(= \supc{\B})$, which we shall call a \textit{consistency} operator, inspired by the Logics of Formal Inconsistency (LFIs) \cite{CM,BookCC16,CCR2020}. From the previous discussion it is easy to see that the LFIs' \textit{principle of gentle explosion} obtains:
$${\bcirc} A \band A \band {\bneg^\C}A \bapprox \bbot \text{\;\;\;i.e.\;}{\bcirc} A\,, A\,, {\bneg^\C}A \vdash \bbot \,.$$
 
Of course, similar results hold in the dual, paracomplete setting. Let us introduce the connective ${\bstar} \equdef \C^\texttt{fp} \;(= \supd{\B})$, which we call a \textit{determinedness} operator, inspired by the Logics of Formal Undeterminedness (LFUs) \cite{Marcos2005,CCR2020}. We can see that the following holds:
 
$${\bstar}A \bprec A \bor {\bneg^\I}A \text{\;\;\;\;i.e.\;}\vdash {\bbstar}A\,, A\,, {\bneg^\I}A \;\;(\text{for\;} {\bbstar}A \equdef \bcmpl\bstar A) \,.$$
 
\begin{figure}
	\centering
	\begin{subfigure}{0.49\textwidth}
		\centering
		\includegraphics[width=.9\textwidth]{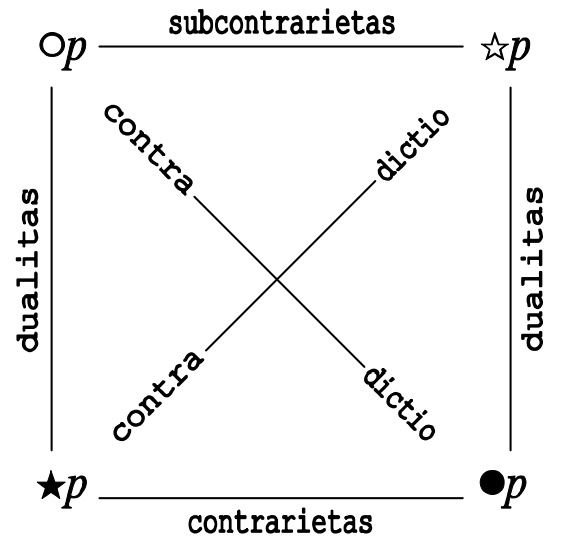}
	\end{subfigure}\hfill%
	\begin{subfigure}{0.49\textwidth}
		\centering
		\includegraphics[width=.9\textwidth]{cube-rear.png}
	\end{subfigure}
	\caption{On the left: Marcos' \textit{square of perfections} \cite{Marcos2005} (edited and amended). On the right: rear face of the \textit{topological cube of opposition} (as in Fig.~\ref{fig:cube-rear}).}
	\label{fig:square-perfections}
\end{figure}

We shall remark that the results stated above (in terms of operators $\C$ and $\I$) have been obtained without assuming any particular restrictions (e.g., Kuratowski conditions). This motivates a more general characterization of recovery operators which will be discussed below. In the meanwhile, before we move the discussion to a more abstract level, it is interesting to recall the \textit{square of perfections} for recovery operators ($\bcirc$ resp.~$\bstar$ and their complements $\bbcirc$ resp.~ $\bbstar$) as introduced by Marcos \cite{Marcos2005} and reproduced slightly amended\footnote{In \cite{Marcos2005} Marcos refers to $\bstar$ as a \textit{determinedness} operator (as we do) but gives it a characterization that is more appropriate for modelling \textit{undeterminedness} (for which we employ $\bbstar$). This issue has also been highlighted by \cite{CCR2020}. As a consequence, the positions for $\bstar$ and $\bbstar$ have been swapped in Fig.~\ref{fig:square-perfections} in contrast to the original diagram as presented by Marcos in \cite{Marcos2005}.}
in Fig.~\ref{fig:square-perfections}, where it is compared to our \textit{topological cube of opposition}.

We now characterize recovery operators in a more abstract fashion. First note that, for any arbitrary operator $\vfi$, if we were to treat $\supdc{\vfi}$ as a negation, then we can see that both ECQ and TND are valid for the fixed points of ${\vfi}$'s dual:\footnote{To see why, the reader can first notice that the set of fixed points of $\supd{\vfi}$ is identical to the set of complements of the fixed points of $\vfi$, i.e., $(\texttt{fp}~\supd{\vfi}) = \supdc{(\texttt{fp}~{\vfi})}$.}
$$(\texttt{fp}~\supd{\vfi})~A \imp A \band \supdc{\vfi}(A) \bapprox \bbot \text{\;\;\;\;\;dually\;\;\;\;\;} (\texttt{fp}~\supd{\vfi})~A \imp A \bor \supdc{\vfi}(A) \bapprox \btop \,.$$

Hence, we can obtain a generalized \textit{principle of gentle explosion} and its dual:
$$\supfp{(\supd{\vfi})}(A) \band A \band \supdc{\vfi}(A) \bapprox \bbot \text{\;\;\;\;\;dually\;\;\;\;\;} \supfp{(\supd{\vfi})}(A) \bprec A \bor \supdc{\vfi}(A) \,,$$
$$\supfp{(\supd{\vfi})}(A)\,, A\,, \supdc{\vfi}(A) \vdash \bbot \text{\;\;\;\;\;dually\;\;\;\;\;} \vdash \supfp{(\supdc{\vfi})}(A)\,, A\,, \supdc{\vfi}(A) \,.$$

Recall from the commutative diagram in Fig.~\ref{fig:cube-abstract} (the abstract \textit{topological cube of opposition}) that $\supfp{(\supd{\vfi})} = \supcfp{(\supdc{\vfi})} $. This gives us another perspective to look at recovery operators. Let us now take an arbitrary negation(-like) operator $\eta$ (playing the role of $\supdc{\vfi}$ above) and state again the generalized \textit{principle of gentle explosion} and its dual:
$$\supcfp{\eta}(A) \band A \band \eta(A) \bapprox \bbot \text{\;\;\;\;\;dually\;\;\;\;\;} \supcfp{\eta}(A) \bprec A \bor \eta(A) \,,$$
$$\supcfp{\eta}(A)\,, A\,, \eta(A) \vdash \bbot \text{\;\;\;\;\;dually\;\;\;\;\;} \vdash \supfp{\eta}(A)\,, A\,, \eta(A) \,.$$

In other words, the transformation $\supcfp{(\cdot)}$ behaves as a `recovery meta-operator' that acts on an operator $\eta$ (intended as a negation) and returns its corresponding recovery operator. In the case of the LFI and LFU operators discussed above, we can see that:

$$\bcirc = \supcfp{({\bneg}^\C)} \text{\;\;\;\;\;dually\;\;\;\;\;} \bstar = \supcfp{({\bneg}^\I)} \;(\text{and so\;} \bbstar = \supfp{({\bneg}^\I)}) \,.$$

Last but not least, we remark that several properties of negation other than ECQ or TND can be recovered by applying fixed-point transformations to topological operators other than $\I$ and $\C$, such as $\B$ or $\F$, thus obtaining other, hitherto unnamed `recovery-like' operators. For instance, the operator $\B^\texttt{fp}(\cdot)$, while not able to recover TND or ECQ, can, under particular conditions, recover some contraposition, De Morgan and double-negation rules. Here again, we refer the reader to \cite{AFP} for more details.

\subsection{Quantifiers}\label{subsec:quantifiers}

We start by recalling from our discussion in \S\ref{subsec:SSE} that quantifiers are introduced in HOL by means of the constants $\Pi^\itype$ of type $(\itype \ar \bool)\ar \bool$. 
They take a `predicate' (i.e.~a term of type $\itype \ar \bool$) and return an appropriate `truth-value' (type $\bool$), namely, \texttt{true} iff the predicate holds (is \texttt{true}) for \textit{all} of the objects in $\itype$'s domain.\footnote{A common definition using equality (and extensionality) is thus: $\Pi^\itype \equdef \lambda P.~(P =^{\itype\ar\bool} \lambda x.~ \texttt{true})$.}
Dually, the constants $\Sigma^\itype$ can be conveniently defined as: $\Sigma^\itype \equdef \lambda P.\,\neg(\Pi^\itype\,(\lambda x.\,\neg(P~x)))$.
We shall recall that the only variable-binding mechanism available in HOL is that of $\lambda$-abstraction, so that the usual variable-binding quantifiers $\forall^\itype$ and $\exists^\itype$ are actually shorthand (`syntactic sugar').
$$\forall^\itype x.~P~x \equdef \Pi^\itype\,(\lambda x.~P) \;\;\;\;\;\;\;\;\;\;\;\; \exists^\itype x.~P~x \equdef \Sigma^\itype\,(\lambda x.~P)$$

In the spirit of shallow semantical embeddings, we introduce for the meta-logical terms $\Pi^\itype$ and $\Sigma^\itype$ their respective $\ww$-type-lifted variants $\bPi^\itype$ and $\bSigma^\itype$ (in \textbf{boldface}) having as type: $(\itype \ar \wsig) \ar \wsig$, i.e., they take a $\ww$-type-lifted predicate (type $\itype \ar \wsig$) and return a $\ww$-type-lifted `truth-value' (type $\wsig$).
$$\bPi^\itype\,\vfi \equdef \lambda w .~\forall^\itype X .~(\vfi~X)~w \;\;\;\;\;\;\;\;\;\;\;\; \bSigma^\itype\,\vfi \equdef \lambda w .~\exists^\itype X .~(\vfi~X)~w$$

We introduce for them convenient variable-binding notation $\bforall^\itype$ and $\bexists^\itype$.%
\footnote{Note the \textbf{boldface}. We omit the superscript $\itype$ in the following. This kind of superscript-notation is also avoided in Isabelle/HOL, where \textit{polymorphically-typed} constants are employed instead.}
$$\bforall X.~\vfi \equdef \bPi\,(\lambda X.~\vfi) \;\;\;\;\;\;\;\;\;\;\;\; \bexists X.~\vfi \equdef \bSigma\,(\lambda X.~\vfi)$$

These quantifiers behave in the expected way, e.g., they are dual to each other.
$$\bPi\,\vfi = \bcmpl(\bSigma\,\supc{\vfi}) \;\;\;\;\;\;\;\;\;\;\;\;\text{i.e.}\;\; (\bforall X.~\vfi~X) = \bcmpl(\bexists X.~\bcmpl(\vfi~X))$$

Thus, using these quantifiers we can write first-order formulas like, e.g., the \textit{drinker's principle} (and actually show it valid automatically).
$$\vdash \bexists x.~\text{Drunk}~x \bimp (\bforall y.~\text{Drunk}~y)$$

Our quantifiers can in fact be given an equivalent formulation in terms of infinitary supremum and infimum operations. Below we employ the notation $\llbracket\cdot\rrbracket$ to denote the range of a function, i.e., $\llbracket f \rrbracket \equdef \lambda y.~\exists x.~(f~x) = y$. Thus, we have that
$$\bPi\,\vfi = \binf\llbracket\vfi\rrbracket \;\;\;\;\;\;\;\;\;\;\;\;\;\;\;\; \bSigma\,\vfi = \bsup\llbracket\vfi\rrbracket$$

or using variable-binder notation
$$\bforall X.~\vfi = \binf \llbracket(\lambda X.~\vfi)\rrbracket \;\;\;\;\;\;\;\; \bexists X.~\vfi = \bsup \llbracket(\lambda X.~\vfi)\rrbracket \,.$$

Observe the similarity to the approach by Mostowski \cite{mostowski1948proofs} of evaluating quantifiers as infinitary lattice operations. In fact, we have seen in \S\ref{subsec:BA} that our encoded Boolean algebras are lattice-complete, so that they naturally lend themselves to this sort of interpretation of quantifiers.\footnote{We can also find this sort of interpretation for quantifiers in Boolean-valued models for set theory; see e.g.~\cite{bell1977boolean}. We think that our approach might also have an application in computer-supported (possibly non-classical) set-theoretical investigations.}

In a similar spirit we can encode quantifiers whose domain of quantification is restricted to a fixed set of objects or individuals. We introduce a pair of quantifiers that take an additional domain set $D$ (of $\itype$-parametric type $(\itype)\sigma$) as a parameter. Hence the new quantifiers will have as a type: $(\itype)\sigma \ar (\itype \ar \wsig) \ar \wsig$.
$$\bPi[D]\,\vfi \equdef \lambda w.~\forall X.~(D~X) \imp (\vfi~X)~w \;\;\;\;\;\; \bSigma[D]\,\vfi \equdef \lambda w.~\exists X.~(D~X) \land (\vfi~X)~w$$

The restricted (or guarded) quantifiers introduced above are said to have a \textit{constant domain}. In this respect they resemble their unconstrained counterparts, where the latter can be regarded as a special case of the former.
$$\bPi\,\vfi = \bPi[\btop]\,\vfi \;\;\;\;\;\;\;\;\;\;\;\;\;\;\;\; \bSigma\,\vfi = \bSigma[\btop]\,\vfi$$

In fact, restricted constant-domain quantifiers can analogously be characterized by means of infima and suprema.\footnote{Recall that we employ the notation $\llbracket\,\cdot~\,\cdot\,\rrbracket$ to denote the image of a set under a function, i.e., $\llbracket f~S \rrbracket \equdef \lambda y.~\exists x.~(S~x)\land~(f~x) = y$.} 
$$\bPi[D]\,\vfi = \binf\llbracket \vfi~D \rrbracket \;\;\;\;\;\;\;\;\;\;\;\;\;\;\;\; \bSigma[D]\,\vfi = \bsup\llbracket \vfi~D \rrbracket$$

Using the equivalences above, and observing that $\llbracket(f\,\circ\,g)~S\rrbracket = \llbracket f~\llbracket g~S\rrbracket \rrbracket$, quantification for function composition can be stated in terms of restricted quantifiers.
$$\bPi(\vfi\,\circ\,\psi) = \bPi[\llbracket \psi \rrbracket]\,\vfi \;\;\;\;\;\;\;\;\;\;\;\;\;\;\;\; \bSigma(\vfi\,\circ\,\psi) = \bSigma[\llbracket \psi \rrbracket]\,\vfi$$

Other, more flexible kind of restricted (or guarded) quantifiers take a set-valued function $\delta(\cdot)$ (we may call it a `domain function'), as an additional parameter of type $\itype \ar \wsig$. This function $\delta$ can be intuitively seen as mapping an ($\itype$-type-)object $X$ to the proposition ``$X$ exists''.\footnote{In line with what some refer to as a (meta-logical) `existence' predicate, which is used to restrict the domains of quantification in presentations of varying-domains semantics for quantified modal logics (cf.~\cite{fitting1998first,fitting2002types} for an exposition).}
Hence varying-domain quantifiers have type $(\itype \ar \wsig) \ar (\itype \ar \wsig) \ar \wsig$.
$$\bPi\{\delta\}\,\vfi \equdef  \lambda w .~\forall X .~(\delta~X)~w \imp (\vfi~X)~w \;\;\;\;\;\;
\bSigma\{\delta\}\,\vfi \equdef \lambda w .~\exists X .~(\delta~X)~w \land (\vfi~X)~w $$

Varying-domain quantifiers are dual to each other in the expected way, i.e., 
$$\bPi\{\delta\}\,\vfi = \bcmpl(\bSigma\{\delta\}\,\supc{\vfi}) \,.$$

It is easily seen that varying-domain quantification generalizes its constant-domain counterpart (below we employ the notation $f{\upharpoonleft} \equdef \lambda x.\,\lambda y.~f\,x$).\footnote{In general $f{\upharpoonleft}$ can be seen as a sort of `inverse projection' that lifts a unary function $f$ to a binary function that is `projected' wrt.~its first argument. This is useful, e.g., for converting a set $D$ into a `rigid' set-valued function $\delta = D{\upharpoonleft}$.}
$$\bPi[D]\,\vfi = \bPi\{D{\upharpoonleft}\}\,\vfi \;\;\;\;\;\;\;\;\;\;\;\;\;\;\;\; \bSigma[D]\,\vfi = \bSigma\{D{\upharpoonleft}\}\,\vfi$$

Recalling that set-valued functions (of type $\itype\ar\wsig$) form a Boolean algebra whose top element is denoted by $\bsqtop$, we can easily present unrestricted quantifiers as a special case of varying-domain quantifiers.
$$\bPi\,\vfi = \bPi\{\bsqtop\}\,\vfi \;\;\;\;\;\;\;\;\;\;\;\;\;\;\;\; \bSigma\,\vfi = \bSigma\{\bsqtop\}\,\vfi$$

In fact, we can employ other Boolean connectives on set-valued functions to encode varying-domain quantifiers as unrestricted ones.

$$\bPi\{\delta\}\,\vfi = \bPi(\delta \,\bsqimp\, \vfi)   \;\;\;\;\;\;\;\;\;\;\;\;\;\;\;\;          \bSigma\{\delta\}\,\vfi = \bSigma(\delta \,\bsqand\, \vfi)$$

In this way, different sorts of restricted quantification can be equivalently encoded using unrestricted variable-binding quantifiers, and employing the connectives $\bsqimp$ and $\bsqand$ to adequately relativize predicates.

$$\bSigma\{\delta\}\,\vfi = \bforall X.~(\delta \,\bsqand\, \vfi)~X \;\;\;\;\;\;\;\;\;\;\; \bPi[D]\,\vfi = \bforall X.~(D{\upharpoonleft} \,\bsqimp\, \vfi)~X$$

\subsubsection{Propositional and Higher-order Quantifiers.}

The reader may have noticed that in our previous discussion we have not specified at which `order' our quantifiers operate, i.e., whether they are intended as first-order, propositional or higher-order quantifiers. In fact, our ($\itype$-parametric) quantifiers behave as first-order quantifiers when instantiating the type parameter $\itype$ with some type for `individuals', say $\iota$. They behave as propositional quantifiers when $\itype$ is instantiated with type $\wsig$. Different sorts of higher-order quantification can be obtaining by instantiating $\itype$ with the appropriate type, e.g., with $\iota \ar \wsig$ for quantifying over (intensional) predicates.\footnote{This makes best sense in the context of an \textit{intensional} higher-order logic; see \cite{fitting2002types} for a thorough discussion and \cite{AFP-Fitting} for its encoding in Isabelle/HOL.}
In the case of, e.g., propositional quantification, the formula below is well-formed (and easily shown valid automatically):
$$\vdash \bforall A.~(\bexists B.~A \bdimp \bcmpl B) \,.$$

An interesting application in non-classical logics arises when we employ quantifiers restricted to domains of fixed points of operators. For the sake of illustration, let us assume below that $\I$ is an operator that satisfies the (duals of) Kuratowski conditions, i.e., $\I$ is a topological interior operator. If we employ quantifiers restricted to the set of $\I$'s fixed points (aka.~open sets) we have that in fact the previous formula does not hold: 
$$\not\vdash \bforall^\I A.~(\bexists^\I B.~A \bdimp \bcmpl B) \,,$$

where $\bforall^\I X.~\vfi = \bPi[\texttt{fp}~\I]\,(\lambda X.~\vfi)$ and $\bexists^\I X.~\vfi = \bSigma[\texttt{fp}~\I]\,(\lambda X.~\vfi)$.

In fact, one can reason in intuitionistic logic using our approach by replacing all (implicit\footnote{For instance, we need to take the universal closure (using $\bforall^\I$) wrt.~all schema/free variables.}) propositional quantifiers in formulas by the ones defined above and employing the paracomplete negation $\bneg^\I \equdef \supdc{\I}$ instead of the classical one (implication needs to be suitably translated\footnote{Since classical implication does not preserve open sets it needs to be suitably encoded, e.g., as done in G\"odel's translation of intuitionistic logic into \textbf{S4}.}). This derives from the  fact that the algebra of open sets of a topological space is a Heyting algebra. The same approach applies \textit{mutatis mutandis} to other non-classical logics with (not necessarily Boolean) lattice-based semantics. They shall be investigated in future follow-up work.

\subsubsection{The Barcan Formula and its Converse}

No discussion of quantification in non-classical contexts would be complete without revisiting the Barcan formula (and its converse).
We start by noting that the converse Barcan formula follows readily from monotonicity for any operator $\vfi$ (wrt.~any arbitrary predicate $\psi$ of type $\itype \ar \wsig$).
\begin{align*}
\MONO~\vfi \imp \vfi\,(\bforall x.~\psi~x) &\bprec \bforall x.~\vfi\,(\psi~x) &\texttt{\;\;\;\;(CBF-1)} \\
\MONO~\vfi \imp \bexists x.~\vfi\,(\psi~x) &\bprec \vfi\,(\bexists x.~\psi~x) &\texttt{\;\;\;\;(CBF-2)}
\end{align*}

So it is in fact the Barcan formula which requires stronger assumptions (of an `infinitary character') in order to hold. 
\begin{align*}
\iMULT^\texttt{b}~\vfi \imp \bforall x.~\vfi\,(\psi~x) &\bprec \vfi\,(\bforall x.~\psi~x) &\texttt{\;\;\;\;(BF-1)}\\
\iADDI^\texttt{a}~\vfi \imp \vfi\,(\bexists x.~\psi~x) &\bprec \bexists x.~\vfi\,(\psi~x) &\texttt{\;\;\;\;(BF-2)}
\end{align*}

Regarding restricted quantification, we observe a difference in the behavior of varying-domain quantifiers in contrast to their constant-domain counterparts. The former generally do not validate the Barcan formula nor its converse, while the latter behave in the same way as their unrestricted counterparts.
For an illustration, let us take the universally-quantified variants for the Barcan formula and its converse, \texttt{BF-1} resp.~\texttt{CBF-1}. First, we repeat the results above for the unrestricted case (recalling that $\MONO~\vfi$ iff $\iMULT^\texttt{a}~\vfi$), but this time without employing variable-binding notation for quantifiers.
\begin{align*}
\iMULT^\texttt{b}~\vfi \imp \bPi\,(\vfi\,\circ\,\psi) &\bprec \vfi\,(\bPi\,\psi) &\texttt{\;\;\;\;(BF-1')} \\
\iMULT^\texttt{a}~\vfi \imp \vfi\,(\bPi\,\psi) &\bprec \bPi\,(\vfi\,\circ\,\psi) &\texttt{\;\;\;\;(CBF-1')}
\end{align*}

The results above hold for quantifiers restricted to constant domains too. The HOL formulas below are thus valid (for an arbitrary domain set $D$ of type $(\itype)\sigma$).
\begin{align*}
\iMULT^\texttt{b}~\vfi \imp \bPi[D]\,(\vfi\,\circ\,\psi) &\bprec \vfi\,(\bPi[D]\,\psi) &\texttt{\;\;\;\;(BF-1-cons)} \\
\iMULT^\texttt{a}~\vfi \imp \vfi\,(\bPi[D]\,\psi) &\bprec \bPi[D]\,(\vfi\,\circ\,\psi) &\texttt{\;\;\;\;(CBF-1-cons)}
\end{align*}

However, in the case of varying domains the results above do not hold. The HOL formulas below are countersatisfiable (for an arbitrary domain function $\delta$ of type $\itype \ar \wsig$). We have obtained countermodels using Isabelle's model generation tool \textit{Nitpick} \cite{Nitpick} even by assuming all (infinitary) interior conditions on $\vfi$.
\begin{align*}
\iMULT\,\vfi \land \CNTR\,\vfi \land \IDEM\,\vfi \not\imp \bPi\{\delta\}\,(\vfi\,\circ\,\psi) &\bprec \vfi(\bPi\{\delta\}\,\psi) &\texttt{\;\;\;\;(BF-1-var)} \\
\iMULT\,\vfi \land \CNTR\,\vfi \land \IDEM\,\vfi \not\imp \vfi(\bPi\{\delta\}\,\psi) &\bprec \bPi\{\delta\}\,(\vfi\,\circ\,\psi) &\texttt{\;\;\;\;(CBF-1-var)}
\end{align*}

The reason why the Barcan formula and its converse hold (under the appropriate conditions) for constant-domain quantifiers (including unrestricted ones) becomes evident if we recall, from the previous subsection, the characterization of constant-domain quantification in terms of infinitary infima and suprema.
$$\bPi[D]\,(\vfi\,\circ\,\psi) = \binf\llbracket (\vfi\,\circ\,\psi)~D \rrbracket \text{\;\;\;and\;\;\;} \bPi[D]\,\psi = \binf\llbracket \psi~D \rrbracket$$
We can now state \texttt{BF-1-cons} and \texttt{CBF-1-cons} somehow more explicitly.
\begin{align*}
\iMULT^\texttt{b}~\vfi \imp \binf\llbracket (\vfi\,\circ\,\psi)~D \rrbracket &\bprec \vfi\,(\binf\llbracket \psi~D \rrbracket) &\texttt{\;\;\;\;(BF-1-cons')} \\
\iMULT^\texttt{a}~\vfi \imp \;\;\vfi\,(\binf\llbracket \psi~D \rrbracket) &\bprec \binf\llbracket (\vfi\,\circ\,\psi)~D \rrbracket &\texttt{\;\;\;\;(CBF-1-cons')}
\end{align*}
Finally, unfolding the definitions for $\iMULT^\texttt{a}$ and $\iMULT^\texttt{b}$ above, and observing that $\llbracket(\vfi\,\circ\,\psi)~D\rrbracket = \llbracket \vfi~\llbracket \psi~D\rrbracket \rrbracket$, we can state \texttt{BF-1-cons} and \texttt{CBF-1-cons} now in a very explicit fashion.
\begin{align*}
(\forall S.~\binf \llbracket \vfi~S \rrbracket \bprec \vfi(\binf S)) \imp \binf\llbracket \vfi~\llbracket \psi~D\rrbracket \rrbracket &\bprec \vfi\,(\binf\llbracket \psi~D \rrbracket) &\texttt{\;\;\;\;(BF-1-cons'')} \\
(\forall S.~\vfi(\binf S) \bprec \binf \llbracket \vfi~S \rrbracket) \imp \;\;\vfi\,(\binf\llbracket \psi~D \rrbracket) &\bprec \binf\llbracket \vfi~\llbracket \psi~D\rrbracket \rrbracket &\texttt{\;\;\;\;(CBF-1-cons'')}
\end{align*}

Thus, we can observe how, in the case of constant-domain quantification, the Barcan formula and its converse become straightforwardly entailed by $\iMULT^\texttt{b}$ and $\iMULT^\texttt{a}$ respectively (or, dually, by $\iADDI^\texttt{a}$ and $\iADDI^\texttt{b}$ respectively). However, the same cannot be said in the context of varying-domain quantification. Loosely speaking, varying-domain quantifiers cannot be `squeezed' into a characterization in terms of infinitary infima and suprema, as in the case of their constant-domain counterparts. They feature an additional `degree of freedom', with their domain of quantification being restricted by a set-valued function rather than just a set.

\section{Conclusion and Prospects}\label{sec:conclusion}

We employed a shallow semantical embedding approach \cite{J41,J23} to encode the key concepts of a broad theory of topological Boolean algebras (cf.~early works by  Kuratowski, Zarycki, McKinsey \& Tarski, among others). We utilized this to provide a natural semantics for \textit{quantified} paraconsistent and paracomplete logics featuring \textit{recovery operators}, particularly certain families of Logics of Formal Inconsistency (LFIs) \cite{CM,BookCC16,CCR2020} and Logics of Formal Undeterminedness (LFUs) \cite{Marcos2005,CCR2020}. We demonstrated how our method facilitates a uniform characterization of propositional, first-order, and higher-order quantification (restricted also to constant and varying domains). The algebraic semantics presented in this paper has, in fact, grown out of an effort to generalize the neighborhood structures introduced in \cite[\S5]{CCF2021}.

A main contribution of our work thus consists in showing how paraconsistent and paracomplete negations, along with recovery operators, can be explicated in terms of (generalized) topological operators, among them modalities (by recalling the notions of Alexandrov topologies and specialization preorders, cf.~\S\ref{subsec:conditions}). The idea of defining non-classical negations by composing a classical negation with a modality is of course not new and has been suggested in numerous occasions by researchers working at the crossroads of modal and non-classical logics. The related papers are too many for us to provide an adequate (and fair) survey. We restrict ourselves to giving a special mention to the works \cite{Marcos2005,Dodo2014} (and references therein) for modal-logic-oriented approaches, and \cite{Baskent2013,CP17} (and references therein) for topologically-oriented approaches relevant to our work. In fact, the latter paper by Coniglio \& Prieto-Sanabria \cite{CP17} is very close in spirit to ours, although the scope of their results differs. Future computer-assisted investigations will strive to bridge their findings with ours. 

An additional contribution involves providing a semantically motivated characterization of quantifiers for the non-classical logics under study. We have begun examining their basic interactions with other connectives, but there is much more to explore.  Leveraging computational tools, we aim to deepen our understanding of the properties of suitably interrelated (combinations of) primitive operators, as exemplified in Rauszer's \textit{bi-topological Boolean algebras} \cite{rauszer1974semi}. We hypothesize that numerous negation(-like) operators examined in the literature can be defined this way (potentially paired with their respective recovery operators). Computer-assisted experiments will be instrumental in shedding further light on their interactions with various types of quantifiers.

So far, the concept of recovery operators has been discussed within the context of paraconsistent and paracomplete logics. Therefore, we have initiated our investigations with the paradigmatic case of negations (and associated modalities). Future work will explore other (binary) connectives. Further generalization of the present theory to provide an unifying semantics for various other systems of non-classical and substructural logics (possibly including relevance and linear logic) is the most likely medium-term outcome of this work. This involves suitably extending the \textit{algebra of operators} presented in \S\ref{subsec:operators}. In the long term, our objective is to provide symbolic ethico-legal reasoning capabilities for AI systems. The present formalization \cite{AFP} complements other works based on the \textit{LogiKEy} framework and methodology \cite{J48} (\url{logikey.org}). Our work could also be of interest to the working logician or mathematician, who might appreciate tedious pen-and-paper calculations being replaced by invocations of model generators (e.g., \textit{Nitpick} \cite{Nitpick}) or automated theorem provers (via \textit{Sledgehammer} \cite{blanchette2016hammering}). Notably, besides their traditional use in formalized mathematics, proof assistants have also been successfully employed in the formal verification of hardware and software systems. One of our goals is thus to apply techniques from formal verification, particularly those using HOL-based systems, to the development of AI systems with logico-pluralistic reasoning capabilities.

Our work has been primarily oriented towards semantics. However, proof-theoretical investigations with the aim of developing sound and complete calculi represent significant potential follow-up work. We are currently exploring techniques to utilize mathematical proof assistants, particularly Isabelle/HOL, for formally verifying proof calculi for non-classical logical systems.  Indeed, fully automatic verification of calculi soundness is straightforward using the shallow semantical embedding approach; it simply requires that we check whether the corresponding rules, encoded as meta-logical implications, follow as theorems from the axiomatic semantic (e.g., frame) conditions plus the definitions of object-logical connectives. Completeness proofs require, however, an explicit encoding of the syntax and rules of the object calculi,\footnote{This corresponds to the notion of a \textit{deep embedding}. We refer the reader to \cite{DeepShallow} for a discussion of the differences (and similarities) between \textit{deep} and \textit{shallow} embeddings.} together with elaborated `quotient' constructions, which demand relatively complex proofs by induction whose full automation remains an open problem \cite{johansson2019lemma}. Assuming, as we do, that our object logics are defined semantically via their shallow embeddings (wrt.~HOL as a meta-language), we only need to allow for digressions into meta-logical reasoning by suitable provers (complete wrt.~HOL \cite{B5,Andreka2014}) during proof search to ensure completeness.

\pagebreak


%
%
%

\begin{thebibliography}{10}
\providecommand{\url}[1]{\texttt{#1}}
\providecommand{\urlprefix}{URL }
\providecommand{\doi}[1]{https://doi.org/#1}

\bibitem{Andreka2014}
Andr{\'e}ka, H., van Benthem, J., Bezhanishvili, N., N{\'e}meti, I.: Changing a
  semantics: Opportunism or courage? In: Manzano, M., Sain, I., Alonso, E.
  (eds.) The Life and Work of Leon Henkin: Essays on His Contributions, pp.
  307--337. Springer International Publishing, Cham (2014)

\bibitem{andrews1972general}
Andrews, P.B.: General models and extensionality. The Journal of Symbolic Logic
   \textbf{37}(2),  395--397 (1972)

\bibitem{andrewsBook}
Andrews, P.B.: An introduction to mathematical logic and type theory: to truth
  through proof, Applied Logic Series, vol.~27. Springer Dordrecht, 2 edn.
  (2002)

\bibitem{Baskent2013}
Başkent, C.: Some topological properties of paraconsistent models. Synthese
  \textbf{190}(18),  4023--4040 (2013)

\bibitem{bell1977boolean}
Bell, J.L.: Set Theory: Boolean-Valued Models and Independence Proofs. Oxford
  Logic Guides, Oxford Scuence Publications (2005)

\bibitem{J41}
Benzm{\"u}ller, C.: Universal (meta-)logical reasoning: Recent successes.
  Science of Computer Programming  \textbf{172},  48--62 (2019)

\bibitem{J6}
Benzm{\"u}ller, C., Brown, C., Kohlhase, M.: Higher-order semantics and
  extensionality. Journal of Symbolic Logic  \textbf{69}(4),  1027--1088 (2004)

\bibitem{B5}
Benzm{\"u}ller, C., Miller, D.: Automation of higher-order logic. In: Gabbay,
  D.M., Siekmann, J.H., Woods, J. (eds.) Handbook of the History of Logic,
  Volume 9 --- Computational Logic, pp. 215--254. North Holland, Elsevier
  (2014)

\bibitem{J48}
Benzm{\"u}ller, C., Parent, X., van~der Torre, L.: Designing normative theories
  for ethical and legal reasoning: {LogiKEy} framework, methodology, and tool
  support. Artificial Intelligence  \textbf{287},  103348 (2020)

\bibitem{J23}
Benzm{\"u}ller, C., Paulson, L.C.: Quantified multimodal logics in simple type
  theory. Logica Universalis (Special Issue on Multimodal Logics)
  \textbf{7}(1),  7--20 (2013)

\bibitem{SEPTT}
Benzmüller, C., Andrews, P.: {Church’s Type Theory}. In: Zalta, E.N. (ed.)
  The {Stanford} Encyclopedia of Philosophy. Metaphysics Research Lab, Stanford
  University, {S}ummer 2019 edn. (2019),
  \url{https://plato.stanford.edu/archives/sum2019/entries/type-theory-church/}

\bibitem{blanchette2016hammering}
Blanchette, J.C., Kaliszyk, C., Paulson, L.C., Urban, J.: Hammering towards
  {QED}. Journal of Formalized Reasoning  \textbf{9}(1),  101--148 (2016)

\bibitem{Nitpick}
Blanchette, J.C., Nipkow, T.: Nitpick: {A} counterexample generator for
  higher-order logic based on a relational model finder. In: Kaufmann, M.,
  Paulson, L.C. (eds.) ITP 2010. LNCS, vol.~6172, pp. 131--146. Springer (2010)

\bibitem{bou2009logics}
Bou, F., Esteva, F., Font, J.M., Gil, {\`A}.J., Godo, L., Torrens, A.,
  Verd{\'u}, V.: Logics preserving degrees of truth from varieties of
  residuated lattices. Journal of Logic and Computation  \textbf{19}(6),
  1031--1069 (2009)

\bibitem{BookCC16}
Carnielli, W.A., Coniglio, M.E.: Paraconsistent Logic: Consistency,
  Contradiction and Negation, Logic, Epistemology, and the Unity of Science,
  vol.~40. Springer, Cham (2016)

\bibitem{CCF2021}
Carnielli, W.A., Coniglio, M.E., Fuenmayor, D.: Logics of formal inconsistency
  enriched with replacement: An algebraic and modal account. The Review of
  Symbolic Logic  \textbf{15}(3),  771–806 (2022)

\bibitem{CCR2020}
Carnielli, W.A., Coniglio, M.E., Rodrigues, A.: Recovery operators,
  paraconsistency and duality. Logic Journal of the IGPL  \textbf{28}(5),
  624--656 (2020)

\bibitem{CM}
Carnielli, W.A., Marcos, J.: A taxonomy of {C}-systems. In: Carnielli, W.A.,
  Coniglio, M.E., D'Ottaviano, I.M.L. (eds.) Paraconsistency: The Logical Way
  to the Inconsistent. Proceedings of the 2nd World Congress on Paraconsistency
  (WCP 2000), Lecture Notes in Pure and Applied Mathematics, vol.~228, pp.
  1--94. Marcel Dekker, New York (2002)

\bibitem{CR2017}
Carnielli, W.A., Rodrigues, A.: An epistemic approach to paraconsistency: a
  logic of evidence and truth. Synthese  \textbf{196}(9),  3789--3813 (2017)

\bibitem{Church40}
Church, A.: A formulation of the simple theory of types. J. Symb. Log.
  \textbf{5}(2),  56--68 (1940)

\bibitem{CP17}
Coniglio, M.E., Prieto-Sanabria, L.: Modal logic {S}4 as a paraconsistent logic
  with a topological semantics. In: P.~Gouveia, C.C., Donisio, F. (eds.) Logic
  and Computation: Essays in Honour of Amilcar Sernadas, Tributes, vol.~33, pp.
  171--196. College Publications, London (2017)

\bibitem{Dodo2014}
Dodó, A., Marcos, J.: Negative modalities, consistency and determinedness.
  Electronic Notes in Theoretical Computer Science  \textbf{300},  21--45
  (2014)

\bibitem{Esakia}
Esakia, L.: Intuitionistic logic and modality via topology. Annals of Pure and
  Applied Logic  \textbf{127}(1-3),  155--170 (2004)

\bibitem{fitting2002types}
Fitting, M.: Types, Tableaus, and G{\"o}del's God, Trends in logic, vol.~12.
  Springer Science \& Business Media (2002)

\bibitem{fitting1998first}
Fitting, M., Mendelsohn, R.L.: First-order modal logic, Synthese Library,
  vol.~277. Springer Science \& Business Media (1998)

\bibitem{AFP}
Fuenmayor, D.: Topological semantics for paraconsistent and paracomplete
  logics. Archive of Formal Proofs  (2020),
  (\sout{\url{https://isa-afp.org/entries/Topological_Semantics.html}}. Check
  instead \url{https://github.com/davfuenmayor/topological-semantics} for an
  up-to-date development version of this entry)

\bibitem{ECAI}
Fuenmayor, D., Benzm{\"u}ller, C.: Normative reasoning with expressive logic
  combinations. In: De~Giacomo, G., Catala, A., Dilkina, B., Milano, M., Barro,
  S., Bugarín, A., Lang, J. (eds.) {ECAI} 2020 -- 24th European Conference on
  Artificial Intelligence, June 8-12, Santiago de Compostela, Spain. Frontiers
  in Artificial Intelligence and Applications, vol.~325, pp. 2903--2904. {IOS}
  Press (2020)

\bibitem{AFP-Fitting}
Fuenmayor, D., Benzmüller, C.: {Types, Tableaus and G\"odel’s God in
  Isabelle/HOL}. Archive of Formal Proofs  (May 2017),
  \url{https://isa-afp.org/entries/Types_Tableaus_and_Goedels_God.html}, Formal
  proof development

\bibitem{CICM}
Fuenmayor, D., Serrano~Su{\'a}rez, F.F.: Formalising basic topology
  for computational logic in simple type theory. In: Buzzard, K., Kutsia, T.
  (eds.) Intelligent Computer Mathematics. pp. 56--74. Springer International
  Publishing, Cham (2022)

\bibitem{DeepShallow}
Gibbons, J., Wu, N.: Folding domain-specific languages: deep and shallow
  embeddings (functional pearl). In: Jeuring, J., Chakravarty, M.M.T. (eds.)
  Proceedings of the 19th {ACM} {SIGPLAN} international conference on
  Functional programming, Gothenburg, Sweden, September 1-3, 2014. pp.
  339--347. {ACM} (2014)

\bibitem{gordon1988hol}
Gordon, M.J.: {HOL}: A proof generating system for higher-order logic. In: VLSI
  specification, verification and synthesis, pp. 73--128. Springer (1988)

\bibitem{gordon1985hol}
Gordon, M.: {HOL} - a machine oriented formulation of higher order logic.
  Technical Report~68, University of Cambridge - Computer Laboratory (July
  1985)

\bibitem{harrison2009hol}
Harrison, J.: {HOL} light: An overview. In: International Conference on Theorem
  Proving in Higher Order Logics. pp. 60--66. Springer (2009)

\bibitem{Hausdorff}
Hausdorff, F.: Grundz{\"u}ge der {M}engenlehre, vol.~7. von Veit (1914)

\bibitem{henkin1950completeness}
Henkin, L.: Completeness in the theory of types. The Journal of Symbolic Logic
  \textbf{15}(2),  81--91 (1950)

\bibitem{johansson2019lemma}
Johansson, M.: Lemma discovery for induction: A survey. In: Intelligent
  Computer Mathematics: 12th International Conference, CICM 2019, Prague, Czech
  Republic, July 8--12, 2019, Proceedings 12. pp. 125--139. Springer (2019)

\bibitem{BAO}
J{\'o}nsson, B., Tarski, A.: Boolean algebras with operators. {Part I}.
  American journal of mathematics  \textbf{73}(4),  891--939 (1951)

\bibitem{Kuratowski1}
Kuratowski, K.: Sur l'op{\'e}ration \={A} de l'analysis situs. Fundamenta
  Mathematicae  \textbf{3}(1),  182--199 (1922)

\bibitem{Kuratowski2}
Kuratowski, K.: Topologie I, vol.~20. Monografie Matematyczne (1948)

\bibitem{Marcos2005}
Marcos, J.: Nearly every normal modal logic is paranormal. Logique et Analyse
  \textbf{48}(189/192),  279--300 (2005)

\bibitem{AOT}
McKinsey, J.C., Tarski, A.: The algebra of topology. Annals of mathematics
  \textbf{45},  141--191 (1944)

\bibitem{mostowski1948proofs}
Mostowski, A.: Proofs of non-deducibility in intuitionistic functional
  calculus. The Journal of Symbolic Logic  \textbf{13}(4),  204--207 (1948)

\bibitem{Isabelle}
Nipkow, T., Paulson, L.C., Wenzel, M.: {Isabelle/HOL}: A Proof Assistant for
  Higher-Order Logic, LNCS, vol.~2283. Springer (2002)

\bibitem{TMMT}
Rasiowa, H., Sikorski, R.: The Mathematics of Metamathematics. Panstwowe
  Wydawnictwo Naukowe (1963)

\bibitem{rauszer1974semi}
Rauszer, C.: Semi-boolean algebras and their applications to intuitionistic
  logic with dual operations. Fundamenta Mathematicae  \textbf{83} (1974)

\bibitem{schoenfinkel1924}
Sch\"onfinkel, M.: {\"Uber die Bausteine der mathematischen Logik}.
  Mathematische Annalen  \textbf{92},  305--316 (1924)

\bibitem{Tarski1933}
Tarski, A.: The concept of truth in the languages of the deductive sciences.
  Prace Towarzystwa Naukowego Warszawskiego, Wydzial III Nauk
  Matematyczno-Fizycznych  \textbf{34}(13-172), ~198 (1933)

\bibitem{Zarycki1}
Zarycki, M.: Quelques notions fondamentales de l'analysis situs au point de vue
  de l'alg{\`e}bre de la logique. Fundamenta Mathematicae  \textbf{9}(1),
  3--15 (1927)

\bibitem{Zarycki2}
Zarycki, M.: Allgemeine {E}igenschaften der cantorschen {K}oh{\"a}renzen.
  Transactions of the American Mathematical Society  \textbf{30}(3),  498--506
  (1928)

\bibitem{Zarycki3}
Zarycki, M.: Some properties of the derived set operation in abstract spaces.
  Nauk. Zap. Ser. Fiz.-Mat.  \textbf{5},  22--33 (1947)

\end{thebibliography}


\end{document}